%% file: bulge.tex
\documentclass[usenatbib,useAMS]{mn2e}
\usepackage{graphics}
\usepackage{psfrag}
\usepackage{xspace}
\usepackage{amstext}
\usepackage{amsmath}
\usepackage{amssymb}
\usepackage{natbib}
\usepackage{aas_macros}
\usepackage{ulem}

\usepackage{times}

\usepackage{graphicx}
\newcommand{\FigDir}[1]{./#1}

\def \kpc {\ensuremath{\rm {kpc}}\xspace}
\def \pc {\ensuremath{\rm {pc}}\xspace}

\def \chisq {\ensuremath{\chi^2}\xspace}

\def \IVmI {\ensuremath{I_{\rm V-I}}\xspace}
\def \NRC {\ensuremath{N_{\rm RC} }\xspace}
\def \IzRC {\ensuremath{I_{0,\rm RC}}\xspace}
\def \VmIzRC {\ensuremath{(V-I)_{0,\rm RC}}\xspace}

\def \sigRC {\ensuremath{\sigma_{\rm RC}}\xspace}
\def \mRC {\ensuremath{I_{V-I, {\rm RC}}}\xspace}

\def \sigLOS {\ensuremath{\sigma_{\rm los}}\xspace}

\def  \sigRCdisp {\ensuremath{\sigma_{\rm RC,0}}\xspace}
\def  \sige {\ensuremath{\sigma_{\rm e}}\xspace}

\def \delmRC {\ensuremath{\xi_{\IVmI,{\rm RC}}} \xspace}
\def \delsigRC {\ensuremath{\xi_{\sigRC}} \xspace}
\def \deg {\ensuremath{^{\circ}}\xspace}

\def \DelIRC {\ensuremath{\Delta I_{\rm RC}}\xspace}

\def \Ro {\ensuremath{R_{0}}\xspace}

\newlength{\voff}
\newcommand{\ppm}{$\pm$}

\begin{document}

\title[Modelling the Galactic bar]{Modelling the Galactic bar using OGLE-II  Red Clump Giant Stars}

\author[Rattenbury et al.]
{Nicholas J. Rattenbury$^{1}$, Shude Mao$^{1}$, Takahiro Sumi$^{2}$, Martin C. Smith$^{3}$ \thanks{e-mail: (njr, smao)@jb.man.ac.uk;  sumi@stelab.nagoya-u.ac.jp; msmith@astro.rug.nl}\\
$^1$ University of Manchester, Jodrell Bank Observatory, Macclesfield, SK11 9DL, UK \\
$^2$ Solar-Terrestrial Environment Laboratory, Nagoya University, Furo-cho, Chikusa-ku, Nagoya, 464-8601, Japan\\
$^3$ Kapteyn Astronomical Institute, University of Groningen, P.O. Box 800, 9700 AV Groningen, The Netherlands
}
\date{Accepted ........
      Received .......;
      in original form ......}

\pubyear{2005}

\maketitle
\begin{abstract}
Red clump giant stars can be used as distance indicators to trace the mass distribution of the Galactic bar. We use RCG stars from 44 bulge fields from the OGLE-II microlensing collaboration database to constrain analytic tri-axial models for the Galactic bar. We find the bar major axis is oriented at an angle of $24^{\circ}$ -- $27^{\circ}$ to the Sun-Galactic centre line-of-sight. The ratio of semi-major and semi-minor bar axis scale lengths in the Galactic plane $x_{0},y_{0}$,  and vertical bar scale length $z_{0}$, is $x_{0}:y_{0}:z_{0} = 10:3.5:2.6$, suggesting a slightly more prolate bar structure than the working model of \citet{2002ASPC..273...73G} which gives the scale length ratios as  $x_{0}:y_{0}:z_{0} = 10:4:3$.
\end{abstract}


\section{Introduction}
\label{sec:intro}

It is now generally accepted that the Galactic bulge is a tri-axial, bar-like structure. Observational evidence for a bar has arisen from several sources, such as the study of gas kinematics (e.g. \citealt{1991MNRAS.252..210B}), surface brightness  (e.g. \citealt{1991ApJ...379..631B}), star counts (e.g. \citealt{1991Natur.353..140N,1994ApJ...429L..73S}) and microlensing (e.g. \citealt{1994AcA....44..165U}); see \citet{2002ASPC..273...73G}  for a review. 

Observational data have been used to constrain dynamical models of the Galaxy. \citet{1995ApJ...445..716D} used the COBE-DIRBE multi-wavelength observations of the Galactic centre \citep{1994ApJ...425L..81W} to constrain several analytic bar models. \citet{1997ApJ...477..163S} used optical observations of red clump giant (RCG) stars to constrain theoretical bar models. Similarly, \citet{2005MNRAS.358.1309B} and \citet{2005ApJ...621L.105N} traced the bulge RCG population in the infrared. This work uses a sample of stars 30 times larger than that of \citet{1997ApJ...477..163S}, with a greater number of fields distributed across a larger area of the Galactic bulge, thus allowing finer constraints to be placed on the bar parameters than those determined by \citet{1997ApJ...477..163S}.

Our current understanding of the Galactic bar is that it is orientated at about $15-40^{\circ}$ to the Sun--Galactic centre line-of-sight, with the near end in the first Galactic longitude quadrant. The bar length is around 3.1 -- 3.5 \kpc with axis ratio approximately $10:4:3$ \citep{2002ASPC..273...73G}. The above bar parameters are generally accepted as a working model, however they are not well determined. Our understanding of the  complete structure of the inner Galactic regions is similarly incomplete.  For example, recent infra-red star counts collected by the Spitzer Space Telescope for Galactic longitudes $l$ = $10\deg$ -- $30\deg$ are best explained assuming a long thin bar oriented at an angle of $\sim 44^\circ$ to the Sun--Galactic centre line \citep{2005ApJ...630L.149B} while most previous studies (performed at $|l| \lesssim 12\deg$) prefer a short bar with an opening angle of  $\sim 20^\circ$. Recently, \citet{2007astro.ph..2109C} report that NIR observations of RCGs support the hypothesis that a long thin bar oriented at $\sim 45^\circ$ co-exists with a distinct short tri-axial bulge structure oriented at $\sim 13^\circ$. In addition, there may be some fine features, such as a ring in the Galactic bulge \citep{2005MNRAS.358.1309B}, or a secondary bar \citep{2005ApJ...621L.105N}, that are not yet firmly established. It is therefore crucial to obtain as many constraints as possible in order to better understand the structure of the inner Galaxy.

In this paper we present an analysis of RCG stars observed in the Galactic bulge fields during the second phase of the OGLE microlensing project \citep{2000AcA....50....1U}. These stars are bright and they are approximately standard candles, hence their magnitudes can be taken as an approximate measure of their distances. Number counts in 34 central bulge fields with $-4^{\circ} \leq l \leq 6^{\circ}$ and $-6^{\circ} \leq b \leq 3^{\circ}$ are used to constrain analytic tri-axial bar models, and thereby obtain estimates on bar parameters. We repeat the analysis with 44 fields with $-6.8^{\circ} \leq l \leq 10.6^{\circ}$. We find the fitted bar parameters support the general orientation and shape of the bar reported by other groups. This paper is organised as follows: in Section~2 we describe the OGLE microlensing experiment and photometry catalogue and we illustrate how RCG stars can be used as approximate distance indicators; in Section~3 we detail how RCGs in the OGLE-II proper motion catalogue are selected; in Section~4 we compute the distance modulus to the red clump in 45 OGLE-II fields and thereby trace the central mass density of the Galaxy; in Section~5 we describe how RCG star count histograms for each field can be used to constrain analytic bar models of the inner Galaxy; our results and their comparison to previous works is given in Section~6 and in Section~7 we discuss the implications and limitations of these results. 

\section{Data}
\label{sec:data}
The OGLE \citep{2000AcA....50....1U} and  MOA \citep{2001MNRAS.327..868B, 2003ApJ...591..204S} microlensing collaborations currently make routine observations of crowded stellar fields towards the Galactic bulge, and issue alerts when a microlensing event is detected. A result of this intense monitoring is the creation of massive photometry databases for stars in the Galactic bulge fields. Such databases are extremely useful for kinematic and population studies of the central regions of the Galaxy.

\begin{figure*}
\psfrag{xlabel}{\hspace{-40pt}\normalsize Galactic longitude $(\deg)$}
\psfrag{ylabel}{\hspace{-35pt}\normalsize Galactic latitude $(\deg)$}

\centering \includegraphics[ height=0.98\hsize, angle=-90]{\FigDir{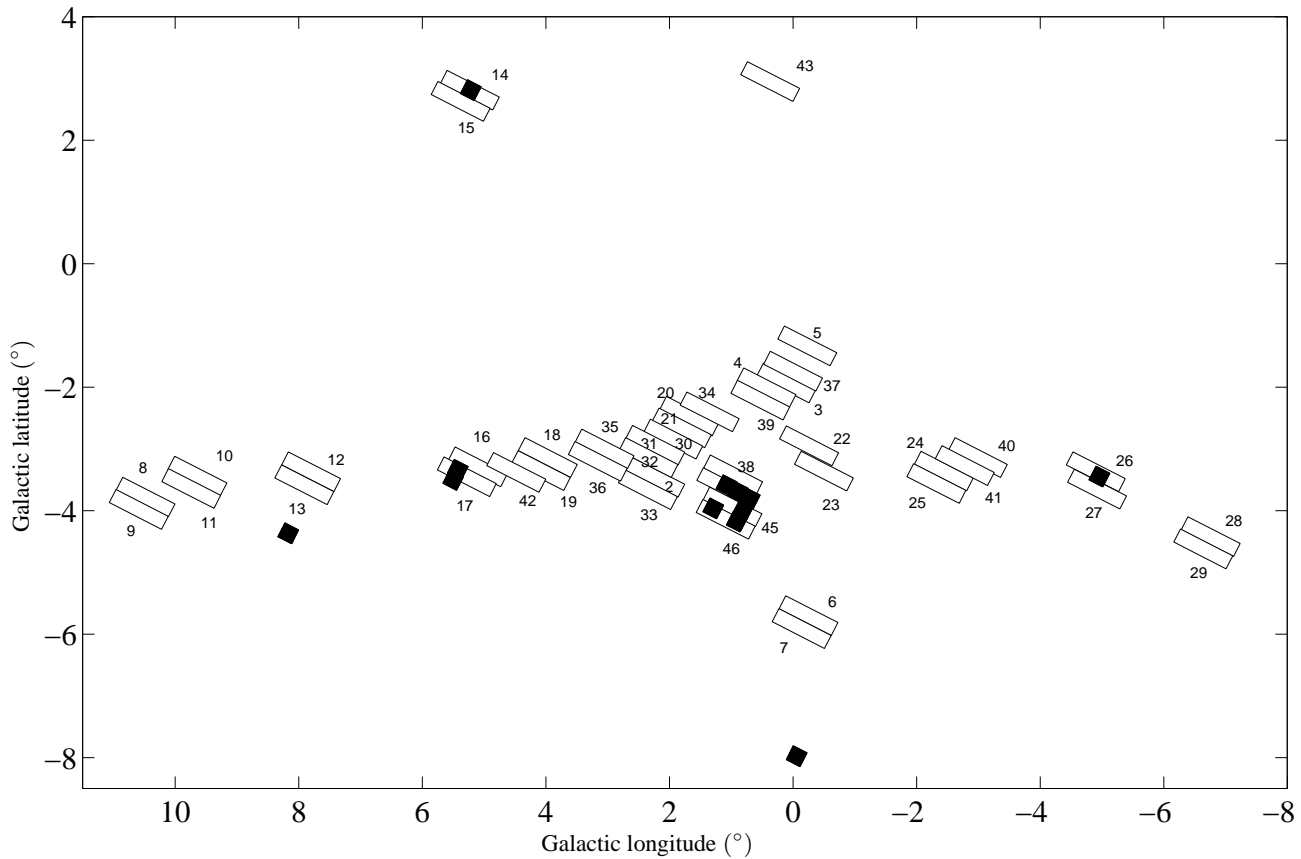}}
\caption{The position of the 45 OGLE-II fields used in this analysis. The solid black regions indicate the location of fields used in \citet{1997ApJ...477..163S}.}
\label{fig:fields}
\end{figure*}

\citet{2004MNRAS.348.1439S} obtained the proper motions for millions of stars in the OGLE-II database for a large area of the sky. Fig.~\ref{fig:fields} shows the OGLE-II fields towards the Galactic bulge. In this paper we focus on the population of red clump giant stars at the Galactic centre. Red clump giants are metal-rich horizontal branch stars \citetext{\citealt{2000AcA....50..191S} and references therein}. Theoretically, one expects their magnitudes to have (small) variations with metallicity, age and initial stellar mass \citep{2001MNRAS.323..109G}. Empirically they appear to be reasonable standard candles in the $I$-band with little dependence on metallicities \citep{2000ApJ...531L..25U, 2001ApJ...551L..85Z}.

\section{Methods}
\citet{1997ApJ...477..163S} used RCG stars in 12 fields (see Fig.~\ref{fig:fields}) observed during the first phase of the OGLE microlensing experiment, OGLE-I, to constrain several analytic models of the Galactic bar density distribution.  \citet{2005MNRAS.358.1309B}, \citet{2005ApJ...621L.105N} and \citet{2007astro.ph..2109C} similarly used IR observations of RCGs to trace the bulge stellar density. We follow similar procedures to extract RCG stars from the OGLE-II Galactic bulge fields and to constrain analytic models.

\subsection{Sample selection}
\label{sec:sample}
\begin{figure}
\psfrag{xlabel}{\raisebox{-5pt} \normalsize $V-I$}
\psfrag{ylabel}{\normalsize $\IVmI$}
\includegraphics[width=1.0\hsize]{\FigDir{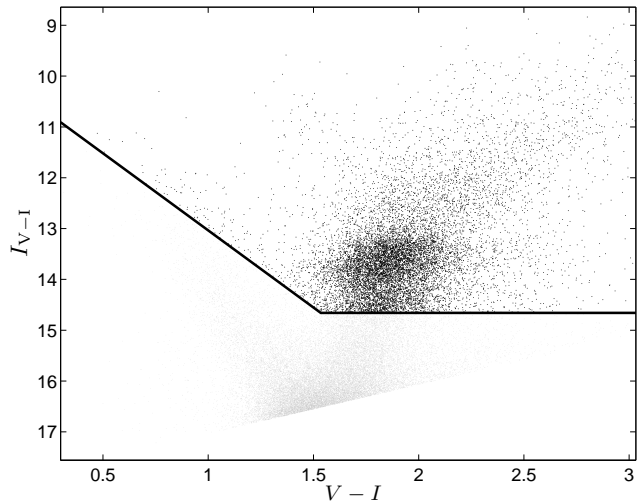}}
\caption{\label{fig:cmd}Reddening-independent magnitude vs colour diagram for OGLE-II field 1. The red clump is clearly visible. Stars are selected (black dots) using the criteria $I < 4(V-I) + k$, where $k$ is a constant chosen for each field, and $\IVmI < 14.66$ (solid lines, see text).}
\end{figure}

We compute the reddening-independent magnitude \IVmI for all stars in each of the 45 OGLE-II fields:
\begin{equation*}
\IVmI = I - A_{\rm I} / (A_{\rm V} - A_{\rm I})\; (V-I)
\end{equation*}
where $A_{\rm I}$ and $A_{\rm V}$ are the extinctions in the $I$ and $V$ bands determined by \citet{2004MNRAS.349..193S}. We select stars which have $I < 4(V-I) + k$, where $k$ is a constant chosen for each field that excludes the main-sequence dwarf stars, and $\IVmI < 14.66$, which corresponds to the magnitude of  RCG stars closer than 15 kpc\footnote{We assume the fiducial RCG star at 15 kpc has an absolute magnitude and colour of $I_{0} = -0.26$ and $(V-I)_{0} = 1.0$ respectively, with $A_{\rm I} / (A_{\rm V} - A_{\rm I}) = 0.96$.}. Fig.~\ref{fig:cmd} shows the sample of stars selected from the $\IVmI,(V-I)$ CMD for OGLE-II field 1.   

\begin{figure}
\psfrag{xlabel}{\raisebox{-5pt} \normalsize $\IVmI$}
\psfrag{ylabel}{\normalsize $N(\IVmI)$}
\includegraphics[width=1.0\hsize]{\FigDir{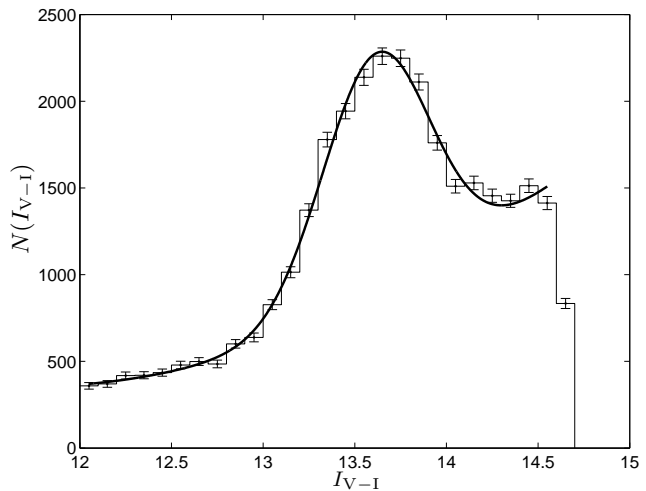}}
\caption{\label{fig:hist}Number count histogram for selected stars in OGLE-II field 1. The heavy solid line is the best-fitting model of the form given by equation~(\ref{eq:fitfun}). The last histogram bin is generally not included in the fitting due to incompleteness effects near the limiting magnitude of $\IVmI < 14.66$.}
\end{figure}

The selected stars are then collected in $\IVmI$ magnitude bins, see Fig.~\ref{fig:hist}. A function comprised of quadratic and Gaussian components is used to model this number count histogram in each of the OGLE-II fields:

\begin{equation}
N(x \equiv \IVmI) = a + bx + cx^{2} + \frac{\NRC}{\sqrt{2\pi}\sigRC} \exp \left[ - \frac{(\mRC - x)^{2}}{2{\sigRC}^{2}} \right]
\label{eq:fitfun}
\end{equation}
where \sigRC is the spread of the red clump giant magnitudes, \mRC is the mean apparent magnitude of the red clump, \NRC and $a$, $b$ and $c$ are coefficients for the Gaussian and quadratic components respectively. These six parameters are allowed to vary for each of the OGLE-II fields and the best-fitting values obtained by minimising $\chisq = \sum_{i=1}^{26}[(N - N_{{\rm obs},i}) / \sigma_{i}]^{2}$ where the sum is taken over all $i = 1\ldots 26$ histogram bins which cover the range $12 \leq \IVmI \leq 14.6$. The error on histogram number counts is assumed to be $\sigma_{i} = \sqrt{N_{{\rm obs},i}}$ , i.e. Poissonian.

The errors on the mean magnitude and distribution width of the red clump stars, \delmRC and \delsigRC respectively, are determined using a maximum-likelihood analysis (see e.g. \citealt{1987AJ.....93.1114L}).

We determine the distance modulus to the red clump in each field as:

\begin{eqnarray}
\mu &=& 5\log(d) - 5 = I - \IzRC \nonumber \\
&=& \mRC + R\,\VmIzRC - \IzRC \label{eq:mu}
\end{eqnarray}
where $d$ is the distance to the red clump measured in parsecs, \mRC is the fitted peak reddening-independent magnitude of the red clump, $R$ is the mean value of  $A_{\rm I} / (A_{\rm V} - A_{\rm I})$ for each field. $\IzRC = -0.26\pm0.03$ \citep{2002ApJ...573L..51A} and  $\VmIzRC = 1.0\pm0.08$ \citep{1998ApJ...494L.219P} are the mean absolute magnitude and colour of the population of local red clump giant stars. We assume that the properties of the local population of red clump giant stars are the same as the Galactic bulge population, however it is likely that  population effects are significant. We discuss the effects of red clump population effects in Section~\ref{sec:popeffects}.

\section{Distance Moduli of RCGs}
\begin{figure}
\psfrag{xlabel}{\normalsize x (\kpc)}
\psfrag{ylabel}{\normalsize y (\kpc)}
\psfrag{A}{\hspace{-15pt} \normalsize $l = -10\deg$}
\psfrag{B}{\hspace{-15pt} \normalsize $l = -5\deg$}
\psfrag{C}{\hspace{-15pt} \normalsize $l = +5\deg$}
\psfrag{D}{\hspace{-15pt} \normalsize $l = +10\deg$}

\includegraphics[width=1.0\hsize]{\FigDir{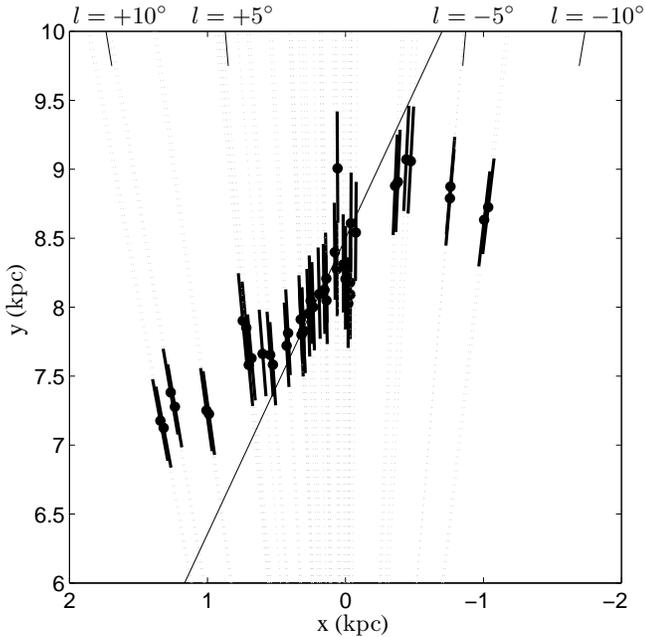}}
\caption{\label{fig:distmod}Structure of the inner region of the Milky Way as traced by red clump giant stars extracted from the OGLE-II microlensing survey data. The mean position of the red clump for 45 OGLE-II fields is indicated by black dots with errors shown by black lines. The grey lines along each line of sight indicate the 1-$\sigma$ spread in distances obtained from fitting equation~(\ref{eq:fitfun}) to the red clump data in each field and correcting for intrinsic red clump luminosity dispersion and photometric errors; see text. A position angle of 25\deg is shown by the thin solid line. Lines-of-sight at $l=\pm 5\deg$ and $l=\pm 10\deg$ are indicated along the top axis.}
\end{figure}

Table~\ref{tab:distmod} lists the fitted parameters $\mRC \pm \delmRC$ and $\sigRC \pm \delsigRC$, along with the mean value of $R$ for each field. The distance modulus, $\mu$, computed from equation~(\ref{eq:mu}) showed that there was a significant offset: the fitted mean red clump giant magnitudes were uniformly too faint compared to that expected of typical local RCG stars at 8 \kpc, resulting in overly large distance moduli. \citet{2004MNRAS.349..193S} found that the OGLE-II RCGs are approximately 0.3 mag fainter than expected when assuming that the population of RCGs in the bulge is the same as local. For this reason, we apply an offset of 0.3 mag to the distance moduli computed above. The shifted distance moduli $\mu' = \mu -0.3$ are given in Table~\ref{tab:distmod}. The true line-of-sight dispersion, \sigLOS, can be approximated by $\sigLOS = (\sigRC^2 - \sigRCdisp^2 - \sige^2)^{1/2}$ where \sigRC is the Gaussian dispersion fitted using equation~(\ref{eq:fitfun}), $\sigRCdisp$  is the intrinsic dispersion of the RCG luminosity function and  $\sige$ is an estimate of the photometric errors  \citep{2005MNRAS.358.1309B}. We use $\sigRCdisp = 0.2$ (see Section~\ref{sec:data}) and $\sige = 0.02$, along with the tabulated values of \sigRC to determine \sigLOS. Fig.~\ref{fig:distmod} shows the mean distance to the red clump stars in each of the 45 OGLE-II fields listed in Table~\ref{tab:distmod}. The fields with Galactic longitude $-4^{\circ} \leq l \leq 6^{\circ}$ show clear evidence of a bar, with a major axis oriented at $\simeq 25^{\circ}$ to the Sun-Galactic centre line-of-sight. For fields with $l > 6^{\circ}$ and $l < -4^{\circ}$ the mean position of the red clump stars do not continue to trace the major axis of a linear structure. \citet{2005MNRAS.358.1309B} find similar evidence that the position of the red clump stars is not predicted by a bar for $l = -9.7^{\circ}$, suggesting that this is a detection of the end of the bar, or the beginning of a ring-like structure. We investigate these possibilities in Section~\ref{sec:ring}. 

The uncertainties on the mean position of the red clump are large; the largest term in the error expression for \delmRC arises from the relatively large uncertainty in the intrinsic colour of the RCGs. The true line-of-sight dispersions, \sigLOS, are consistent with a wide range of bar position angles, but the mean position of the red clump in each direction strongly suggest a bar oriented along a direction consistent with that determined by the previous work referred to in Section~\ref{sec:intro}.

\begin{table*}
\caption{\label{tab:distmod}Fitted values of the red clump mean magnitude, \mRC, and Gaussian dispersion, \sigRC for RCG stars selected from 45 OGLE-II fields. The  mean selective extinction $R$ is also given. The shifted distance modulus  $\mu' = \mu -0.3$ for each field is computed via equation~(\ref{eq:mu}), see text. \sigLOS is the true line-of-sight distance dispersion of the red clump giant stars. $N$ is the total number of stars selected from each CMD, see Section~\ref{sec:sample}}
\begin{tabular}{ccccccccc}
\hline
Field & $l$ & $b$ & \mRC &  \sigRC & $R$ & $\mu'$ & \sigLOS & $N$ \\
\hline
\input{DisModtexTable.txt}
\hline
\end{tabular}
\end{table*}

\subsection{RCG population effects}
\label{sec:popeffects}

\citet{2004MNRAS.349..193S} found that the extinction-corrected magnitudes of RCG stars in the OGLE-II bulge fields were 0.3 mag higher than that of a RCG with an intrinsic magnitude equal to those of local RCGs, placed at a distance of 8\kpc. \citet{2004MNRAS.349..193S} notes that the cause of this offset is uncertain but may be resolved when detailed V-band OGLE-II photometry of RR Lyrae stars will allow improved extinction zero-point estimations for all fields. There is evidence that the properties of bulge RCGs are different to local RCG stars, in particular age and metallicity, resulting in a different average absolute magnitude (see e.g. \citet{2003MNRAS.343..539P,2003ApJ...588..801S}). This could in part explain the observed offset of 0.3 mag between local and bulge RCG stars. The distance modulus plotted in Fig.~\ref{fig:distmod} is as for equation~(\ref{eq:mu}):
\begin{equation*}
\mu = \mRC + R\,\VmIzRC - \IzRC + \kappa
\end{equation*}
where $\kappa = -0.3$ mag is the offset between the observed bulge RCG population and a fiducial local RCG placed at 8 \kpc. Including the difference in intrinsic magnitude between local and bulge RCG populations we have:
\begin{equation*}
\mu = \mRC + R\,\VmIzRC - \IzRC + \DelIRC + \nu
\end{equation*}
where \DelIRC is the intrinsic RCG magnitude difference between local and bulge populations and $\nu$ is a magnitude offset arising from effects other than population differences. The theoretical population models of \citet{2001MNRAS.323..109G} estimate $\DelIRC \simeq -0.1$. The remaining magnitude offset $\nu = -0.2$ can be accounted for by decreasing the assumed value of the Galactocentric distance \Ro from 8 \kpc to 7.3 \kpc. This assumption of \Ro is in agreement with the value of $7.52 \pm 0.10$ (stat) $\pm 0.32$ (sys) \kpc determined by \citet{2006ApJ...647.1093N}, and that of $7.63\pm 0.32$ \kpc determined by \citet{2005ApJ...628..246E}. We note that as long as the stellar population is uniform in the Galactic bulge (an assumption largely consistent with the data), then this unknown offset only affects the zero point (and hence the distance to the Galactic centre), and the absolute bar scale lengths. However the ratios between bar scale lengths are very robust, which we demonstrate in Section~\ref{sec:changeR0} where we determine the bar parameters for several values of \Ro.

A metallicity gradient across the Galactic bulge would have an effect on the intrinsic colour and absolute magnitude of bulge RCG stars as a function of Galactic longitude. \citet{Per06} find that there is a connection between  metallicity gradient and structural features in a sample of 6 barred galaxies. \citet{1995MNRAS.277.1293M} measured metallicities for K giants in two fields at 1.5 \kpc and 1.7 \kpc from the Galactic centre. The average metallicity was [Fe/H] = $-0.6$, lower than that of K giants in Baade's window. However, no such metallicity gradient was reported  for Galactocentric distance range 500 pc -- 3.5 \kpc by \citet{1995MNRAS.275..605I}. Similarly, \citet{2006A&A...458..113S} determine that there is no metallicity gradient within angles $2.2^{\circ}$ to $6.0^{\circ}$ of the Galactic centre (corresponding to 300 -- 800 pc from the GC assuming a Sun-GC distance of 8 \kpc).

\section{Modelling the bar}
\label{sec:modelling}
We continue to investigate the structure of the Galactic bar by fitting analytic models of the stellar density to the observed RCG data in the OGLE-II fields. Following \citet{1997ApJ...477..163S} we use the analytic models of \citet{1995ApJ...445..716D} to fit the observed data. Three model families (Gaussian, exponential and power-law) are used:
{\allowdisplaybreaks
\begin{align}
\rho_{\rm G1} & =  \rho_{0} \exp(-r^{2} / 2) \label{eq:G1}\\
\rho_{\rm G2} & =  \rho_{0} \exp(-r_{\rm s}^{2} / 2)  \\
\rho_{\rm G3} & =  \rho_{0} r^{-1.8}\exp(-r^{3}) \\[20pt]
\rho_{\rm E1} & =  \rho_{0} \exp(-r_{\rm e}) \\
\rho_{\rm E2} & =  \rho_{0} \exp(-r)  \label{eq:E2} \\
\rho_{\rm E3} & =  \rho_{0} K_{0}(r_{\rm s}) \\[20pt]
\rho_{\rm P1} & =  \rho_{0} \left(\frac{1}{1+r}\right)^{4}\\
\rho_{\rm P2} & =  \rho_{0} \left(\frac{1}{r(1+r)^{3}}\right) \\
\rho_{\rm P3} & =  \rho_{0} \left(\frac{1}{1+r^{2}}\right)^{2} \label{eq:P3} \\[10pt] 
\notag
\end{align}
where $K_{0}$ is the modified Bessel function of the second kind and

\begin{align*}
r & \equiv \left[\left(\frac{x}{x_{0}}\right)^{2} + 
                  \left(\frac{y}{y_{0}}\right)^{2} + 
                  \left(\frac{z}{z_{0}}\right)^{2}\right]^{1/2} \\
r_{\rm e} & \equiv  \left[\frac{|x|}{x_{0}} + 
                        \frac{|y|}{y_{0}} + 
                        \frac{|z|}{z_{0}} \right] \\
r_{\rm s} & \equiv \left[\left[\left(\frac{x}{x_{0}}\right)^{2} +
                               \left(\frac{y}{y_{0}}\right)^{2}\right]^{2} +
                        \left(\frac{z}{z_{0}}\right)^{4} \right]^{1/4}
\end{align*} }

The co-ordinate system has the origin at the Galactic centre, with the $xy$ plane defining the mid-plane of the Galaxy and the $z$ direction parallel to the direction of the Galactic poles. The $x$ direction defines the semi-major axis of the bar. The functions are rotated by an angle $\alpha$ around the $z$-axis. An angle of $\alpha = \pm \frac{\pi}{2}$ corresponds to the major axis of the bar pointing towards the Sun. The functions can also be rotated by an angle $\beta$ around the $y$ axis, corresponding to the Sun's position away from the mid-plane of the Galaxy.

We aim to fit the observed number count histograms for each field as a function of magnitude. Given a magnitude \IVmI, the number of stars with this magnitude is \citep{1997ApJ...477..163S}:
\begin{equation}
\label{eq:numcount}
N(\IVmI) = c_{1} \int^{s_{\rm max}}_{s_{\rm min}} \rho(s) s^{2} \Phi(L)L ds
\end{equation}
where the integration is taken over distance $s_{\rm min} = 3\,\kpc < s < s_{\rm max} = 13\,\kpc$. We perform the integration over this range of $\Ro \pm 5$ \kpc as we do not expect the tri-axial bulge density structure to exceed these limits. The constant $c_{1}$ is dependent on the solid angle subtended around each line of sight. The luminosity $L$ is given by
\begin{equation*}
L = c_{2} s^{2} 10^{-0.4\IVmI}
\end{equation*} 
and $c_{2}$ is a constant. The luminosity function $\Phi(L)$ is 
\begin{equation*}
\Phi(L) = N_{0} \left(\frac{L}{L_\odot}\right)^{-\gamma} + \frac{N_{\rm RC}}{\sqrt{2\pi}\,\sigma_{\rm RC}} \exp\left[- \frac{(L - L_{\rm RC})^{2}}{2\sigma_{\rm RC}^{2}}\right]
\end{equation*}
where $L_{\rm RC}$ is the luminosity of the red clump and $\sigma_{\rm RC}$ is the intrinsic spread in red clump giant luminosity and is held constant.

There are ten parameters to be determined in the above equations: the three bar scale lengths $x_{0}$,  $y_{0}$,  $z_{0}$; the bar orientation and tilt angles $\alpha$ and $\beta$; the luminosity function parameters $N_{0}$, $N_{\rm RC}$, $\gamma$ and $L_{\rm RC}$ and the density function parameter $\rho_{0}$. 

There is evidence that the centroid of the bar is offset from the centre of mass of the Galaxy, a feature commonly observed in external galaxies (\citealt{1997ApJ...477..163S,2006ApJ...647.1093N} and references therein). We include an additional parameter in the modelling process, $\delta l$, which allows for this possible centroid offset. Theoretical number counts are computed for the $i=1 \ldots M$ fields at longitudes $l_{i}$ as $N_{i}(l_{i}+\delta l)$ where the offset parameter $\delta l$ is determined over all fields for a given density model.

We apply an exponential cut-off to the density functions similar to that of \citet{1995ApJ...445..716D}:

\begin{equation}
f(r)  = 
        \begin{cases}
        1.0 & r = \left(x^2 + y^2\right)^{1/2} \leq r_{\rm max} \\
        \exp\left(-\frac{(r-r_{\rm max} )^2}{2r_{0}^2}\right) & r > r_{\rm max}  
        \end{cases}
\label{eq:cutoff}
\end{equation}
where $r$ is in kpc, $r_0 = 0.5\,\kpc$ and $r_{\rm max}$ is a cut-off radius. The theoretical constraint on the maximum radius of stable stellar orbits is the co-rotation radius, $r_{\rm max}$. We adopt the co-rotation value of $r_{\rm max} = 2.4\,\kpc$ determined by  \citet{1991MNRAS.252..210B} in equation~(\ref{eq:cutoff}). We later repeat the modelling process without this theoretical cut-off, see Section~\ref{sec:changeR0}.

\citet{1997ApJ...477..163S} included a further fitting parameter to account for a possible metallicity gradient across the bulge, but found that this did not significantly affect the bar parameters. The discussion in Section~\ref{sec:popeffects} suggests that there is no appreciable metallicity gradient over the bulge region investigated here. We therefore do not include an additional model parameter corresponding to a metallicity gradient in the model fitting analysis. 

The model fitting was performed using a standard non-linear Neadler-Mead minimisation algorithm. For each of the nine density profile models we minimise $\chisq$:
\begin{equation*}
\chisq = \sum_{k=1}^{M} \sum_{i=1}^{26} \frac{(N_{ik}(\IVmI) - \widehat{N}_{ik}(\IVmI))^2}{\sigma_{ik}^2}
\end{equation*}
where the summations are taken over  each of the 26 \IVmI histogram bins  in $M$ OGLE-II fields, $N_{k}(\IVmI)$ is the observed histogram data for field $k$ and $\widehat{N}_{k}(\IVmI)$ is the model number count histogram from equation~(\ref{eq:numcount}). The error in $N_{ik}(\IVmI)$ was taken to be $\sigma_{ik} = \sqrt{N_{ik}(\IVmI)}$.

\section{Results}
\label{sec:results}
The 11 parameters $x_{0}$,  $y_{0}$, $z_{0}$, $\alpha$, $\beta$, $A$, $N_{0}$, $N_{\rm RC}$, $\gamma$, $L_{\rm RC}$ and $\delta l$ were fitted for each of the nine density profiles given in equations.~(\ref{eq:G1} -- \ref{eq:P3}) for the 34 OGLE-II fields\footnote{Field 5 was excluded due to poor understanding of the dust extinction in this field.} which have $-4^{\circ} \leq l \leq 6^{\circ}$. A naive interpretation of Fig.~\ref{fig:distmod} is that fitting all fields including those with $l > 6^{\circ}$ and $l < -4^{\circ}$ with a single bar-like structure would not be successful and indeed initial modelling using data from all fields resulted in bar angles of $\simeq 45^{\circ}$. As seen in Fig.~\ref{fig:distmod} this result is consistent with the mean position angle of all fields, but clearly does not describe the bar correctly. We therefore begin modelling the bar using data from the central 34 fields which have $-4^{\circ} \leq l \leq 6^{\circ}$. In Section~\ref{sec:ring} we proceed by including the data from fields with  $l > 6^{\circ}$ and $l < -4^{\circ}$ in the modelling process and in Section~\ref{sec:changeR0} we consider the effect of changing \Ro.

The range of field latitudes is small and therefore the information available in the latitude direction from the total dataset unlikely to be able to constrain  $\beta$ strongly. Initial modelling runs held  $\beta$ constant at $0^{\circ}$. Once a \chisq minimum was determined for each model holding $\beta = 0^{\circ}$, this parameter was allowed to vary with the ten other parameters. 

\begin{figure*}
\psfrag{xlabel}{\normalsize \raisebox{-5pt} \IVmI}
\psfrag{ylabel}{\normalsize \raisebox{5pt} {$N(\IVmI)$}}
\includegraphics[height=0.95\vsize]{\FigDir{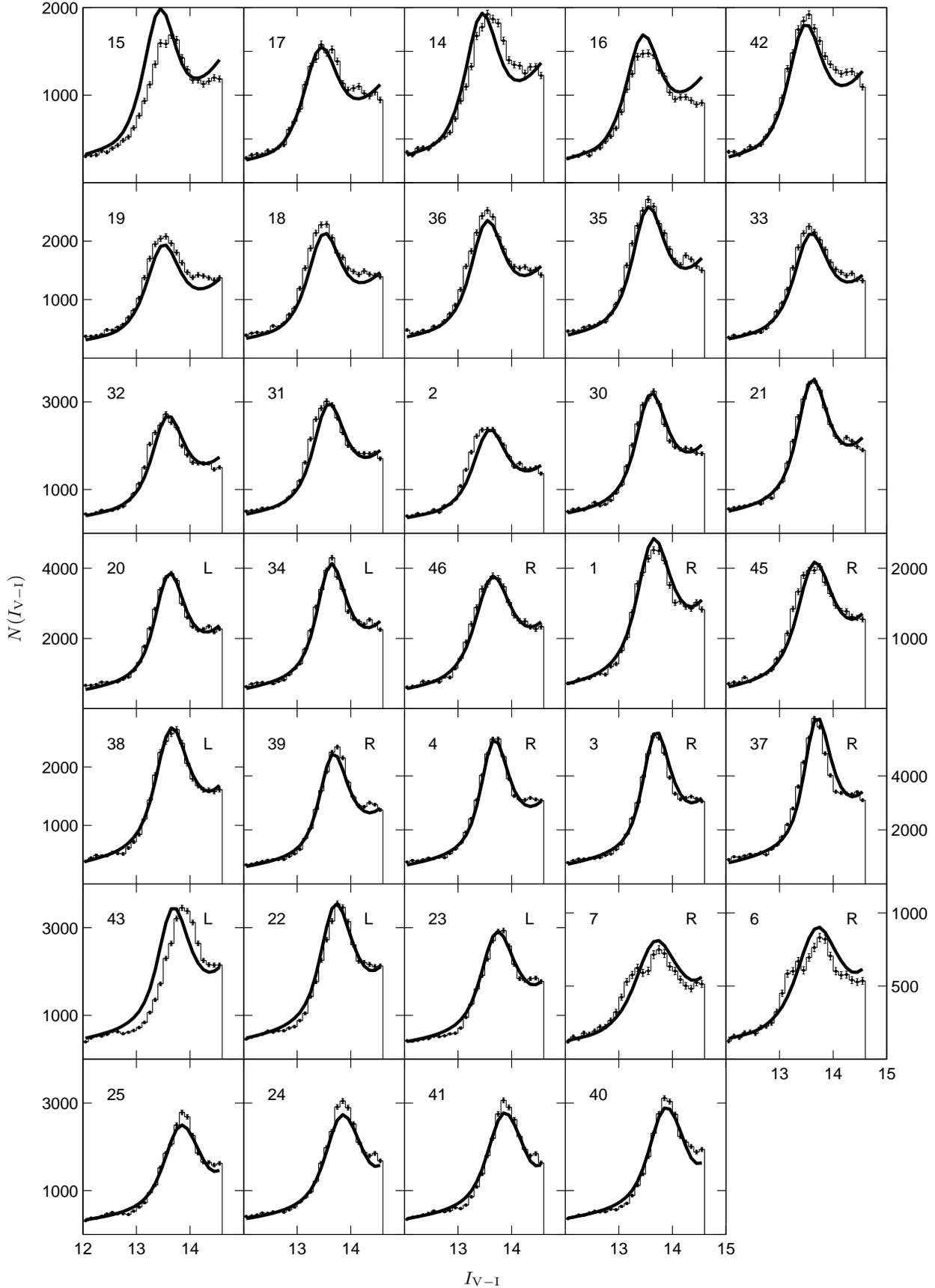}}
\caption{\label{fig:matrixE2}Red clump number count histograms and best-fitting profiles using density model E2 for 34 OGLE-II fields with $-4^{\circ} \leq l \leq 6^{\circ}$. Fields are arranged roughly in order of descending Galactic longitude. Field numbers are given in each set of axes. A `L' or `R' in the axes of  rows 4 -- 6 indicates whether the vertical scale corresponds to left or right vertical axis of the row.}
\end{figure*}

Figure \ref{fig:matrixE2} shows the number count histograms of observed red clump giants in the 34 OGLE-II fields with $-4^{\circ} \leq l \leq 6^{\circ}$, along with the best-fitting model (equation~(\ref{eq:numcount})) using the E2 density profile given in equation~(\ref{eq:E2}) as an example\footnote{Similar figures showing the best-fitting models to the data using all density profiles equations~(\ref{eq:G1} -- \ref{eq:P3}) appear in the on-line supplementary material.}. Each set of axes is arranged roughly in order of decreasing Galactic longitude. The magnitude of the red clump peak increases with decreasing longitude, consistent with a bar potential. 





The observed number count histograms are reasonably well-fitted by all models in most fields. The most obvious exceptions are fields 6, 7, 14, 15 and 43. These fields are at the most extreme latitudes represented in the data. In the case of fields 14 and 15 the algorithm fails to fit satisfactorily both observed histograms for a given model. Upon closer inspection of the observed histogram data for these two fields, it is noted that the total number of stars in the field 15 histogram is significantly less than that for field 14. The models preferred by all fields clearly cannot reproduce the decline in total star count between fields 14 and 15. Upon inspection of the CCD pixel position of stars for field 15 it was found that there are no stars recorded in a region along one edge of the field. The lack of stars in this strip is due to missing $V$-band data for these stars. Similarly, there is a lack of stars in field 15 in the lower right corner, due to heavy dust extinction in this area. 

All models predict the peak of the red clump to be at a lower magnitude than that observed for field 43. A similar offset is seen in the other high latitude fields 14 and 15. These fields are all at latitude $b\simeq3^{\circ}$ and the observed offset in all three fields is likely to be related to this common position in latitude. The cause of this effect is currently unknown. The observed offset between the maximum density position predicted by the tri-axial bar models and that observed for some fields may imply some asymmetry of the bar in the latitude direction, as suggested by \citet{2006ApJ...647.1093N} and references therein.

Fields 6 and 7, both at $b\simeq-6^{\circ}$, are also poorly fitted by all models. The histograms of observed red clump stars in these fields show a curious double peak. Clearly the density functions equations~(\ref{eq:G1} -- \ref{eq:P3}) are inadequate for describing such a feature. This double-peaked structure may be the result of another population of stars lying along the line of sight, at a distance different to the main bulge population. The best-fitting model curves typically have a peak at magnitudes corresponding to the highest peak in the observed histogram, thereby suggesting that the second population, if real, and composed of a significant fraction of RCGs, exists between us and the bulge population.  A population of asymptotic giant branch (AGB) stars \citep{1999ApJ...511..225A,2007AJ....133.1275F} may also be the cause of the secondary peak in the number count histograms for these and similar fields (e.g. field 10, Fig.~\ref{fig:histMatrixWide_G}). Closer investigation of the stellar characteristics and kinematics of stars in these fields may help in describing the origin of the second number count peak.

The best-fitting parameter values for all the nine density models are presented in Table~\ref{tab:paramVals}. Two sets of parameter values are given, one where the tilt angle, $\beta$, was held at $0.0^{\circ}$, and the other where $\beta$ was allowed to vary. The values of \chisq are well in excess of the number of degrees of freedom: $N_{\rm dof} = 26\times34 -11 = 873$, and can therefore not be used to set reliable errors on the parameters values in the usual way. It is likely that the errors on the histogram data are underestimated, resulting in  extreme values of \chisq. We assumed that the errors on the number count data would be Poissonian and we have not added any error in quadrature that might arise from any systematic effects.  Furthermore, as mentioned above, the extreme latitude fields 6, 7, 14, 15 and 43 are poorly fit by every model. The cause for this is unknown, however the effect is to add a relatively high contribution to the total value of \chisq compared to other fields. 

The values of \chisq can only be used to differentiate between the various models in a qualitative manner. In terms of relative performance, model E3  best reproduces the observed number count histograms for the fields tested here, and model G2 provides the worst fit to the data. 

\begin{table*}
\caption{\label{tab:paramVals}Best fitting parameter values for all density models (equations~\ref{eq:G1} -- \ref{eq:P3}) fitted to the number count histograms from the 34 OGLE-II fields with $-4^{\circ} \leq l \leq 6^{\circ}$. Two sets of parameter values are given, one where the tilt angle, $\beta$, was held at $0.0^{\circ}$, and the other where $\beta$ was allowed to vary. $\alpha'$ is the position angle of the bar semi-major axis with respect to the Sun-Galactic centre line-of-sight.}
\begin{tabular}{cccccccc}
\hline
& \multicolumn{2}{c}{Bar orientation $(^{\circ})$} & \multicolumn{3}{c}{Scale lengths (\pc)} & & Axis ratios  \\ 
Model & $\beta$ &  $\alpha'$ & $x_{0}$ & $y_{0}$ & $z_{0}$   & \chisq & $x_{0}:y_{0}:z_{0}$ \\
\hline
\input{paramTexTable.txt}
\hline
\end{tabular}
\end{table*}

The position angle of the bar semi-major axis with respect to the Sun-Galactic centre line-of-sight is $\alpha'$ and is related to the rotation angle $\alpha$ by $\alpha' = \alpha - \rm{sgn}(\alpha)\pi/2$. We find $\alpha' = 20^{\circ}$ -- $26^{\circ}$. The addition of $\beta$ as another variable parameter does not result in a significant improvement in \chisq. The lack of information in the latitude direction means that the data have little power in constraining parameters such as $\beta$.

\begin{figure}
\psfrag{xlabel}{\normalsize Model}
\psfrag{ylabel}{\normalsize Axis Ratio}
\psfrag{XXXXXXXXXXX}{\small $x_{0}/y_{0}$, 34 fields}
\psfrag{second}{\small  $x_{0}/z_{0}$, 34 fields}
\psfrag{third}{\small $x_{0}/y_{0}$, 44 fields}
\psfrag{fourth}{\small $x_{0}/z_{0}$, 44 fields}

\includegraphics[width=1.0\hsize]{\FigDir{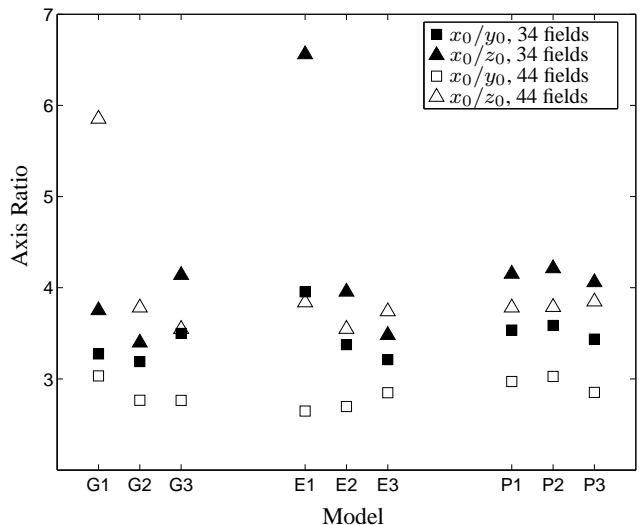}}
\caption{\label{fig:axisRatios}Scale length ratios  $x_{0}/y_{0}$ (squares) and $x_{0}/z_{0}$ (triangles) for all models. Solid and open symbols show the ratios for the best-fitting models using the 34 fields with  $-4^{\circ} \leq l \leq 6^{\circ}$, and all 44 fields (see Section~\ref{sec:ring}) respectively.}
\end{figure}

The absolute values of the scale lengths in Table~\ref{tab:paramVals} cannot be directly compared between models. The axis ratios $x_{0}/y_{0}$ and $x_{0}/z_{0}$ can be compared however. Fig.~\ref{fig:axisRatios} shows the axis ratios  $x_{0}/y_{0}$ and $x_{0}/z_{0}$ for the nine models tested. The ratio of the major bar axis scale length to the minor bar axis in the plane of the Galaxy, $x_{0}/y_{0}$, has values $3.2$ -- $3.6$, with the exception of that for model E1, for which $x_{0}/y_{0} = 4.0$. \citet{1997ApJ...477..163S} found $x_{0}/y_{0} = 2.0$ -- $2.4$, with the exception of model E1 for which  $x_{0}/y_{0} = 2.9$. It is interesting to note that our values of $x_{0}/y_{0}$  have a similar range to that of \citet{1997ApJ...477..163S}, when we exclude the outlying result from the same model (E1). Our results do suggest a slimmer bar, i.e. higher values of $x_{0}/y_{0}$ compared to \citet{1997ApJ...477..163S}. Both these results and the results of \citet{1997ApJ...477..163S} are consistent with the value of $x_{0}/y_{0} \simeq 3\pm 1$ reported by \citet{1995ApJ...445..716D}.

 The ratio of the major bar axis scale length to the vertical axis scale length $x_{0}/z_{0}$ was found to be $3.4$ -- $4.2$, with the same exception of model E1, which has an outlying value of $x_{0}/z_{0} = 6.6$. Again, the range of ratio values is comparable to that of \citet{1997ApJ...477..163S} who found $x_{0}/z_{0} = 2.8$ -- $3.8$, with the exception of model E1 again, which had $x_{0}/z_{0} = 5.6$.

The mean axis ratios are $x_{0}:y_{0}:z_{0}$ are $10$ : $2.9$ : $2.5$. Excluding the outlying results from the E1 model, the ratios are  $10$ : $3.0$ : $2.6$. These results suggest a bar more prolate than the general working model with $10:4:3$ \citep{2002ASPC..273...73G}.

\subsection{Evidence of non-bar structure? Including wide longitude fields}
\label{sec:ring}

Fig.~\ref{fig:distmod} shows the mean position of the bulge red clump stars. At longitudes $-4^{\circ} \leq l \leq 6^{\circ}$ the bulge red clump stars follow the main axis of the bar. At greater angular distances, the mean positions of the red clump stars are clearly separated from the main bar axis. \citet{2005MNRAS.358.1309B} find similar evidence and suggest that this could indicate either the end of the bar or the existence of a ring-like structure. We used the OGLE-II data to investigate these possibilities. 

\begin{figure*}
\psfrag{xlabel}{\normalsize \raisebox{-5pt} \IVmI}
\psfrag{ylabel}{\normalsize \raisebox{5pt} {$N(\IVmI)$}}
\includegraphics[height=0.95\textheight]{\FigDir{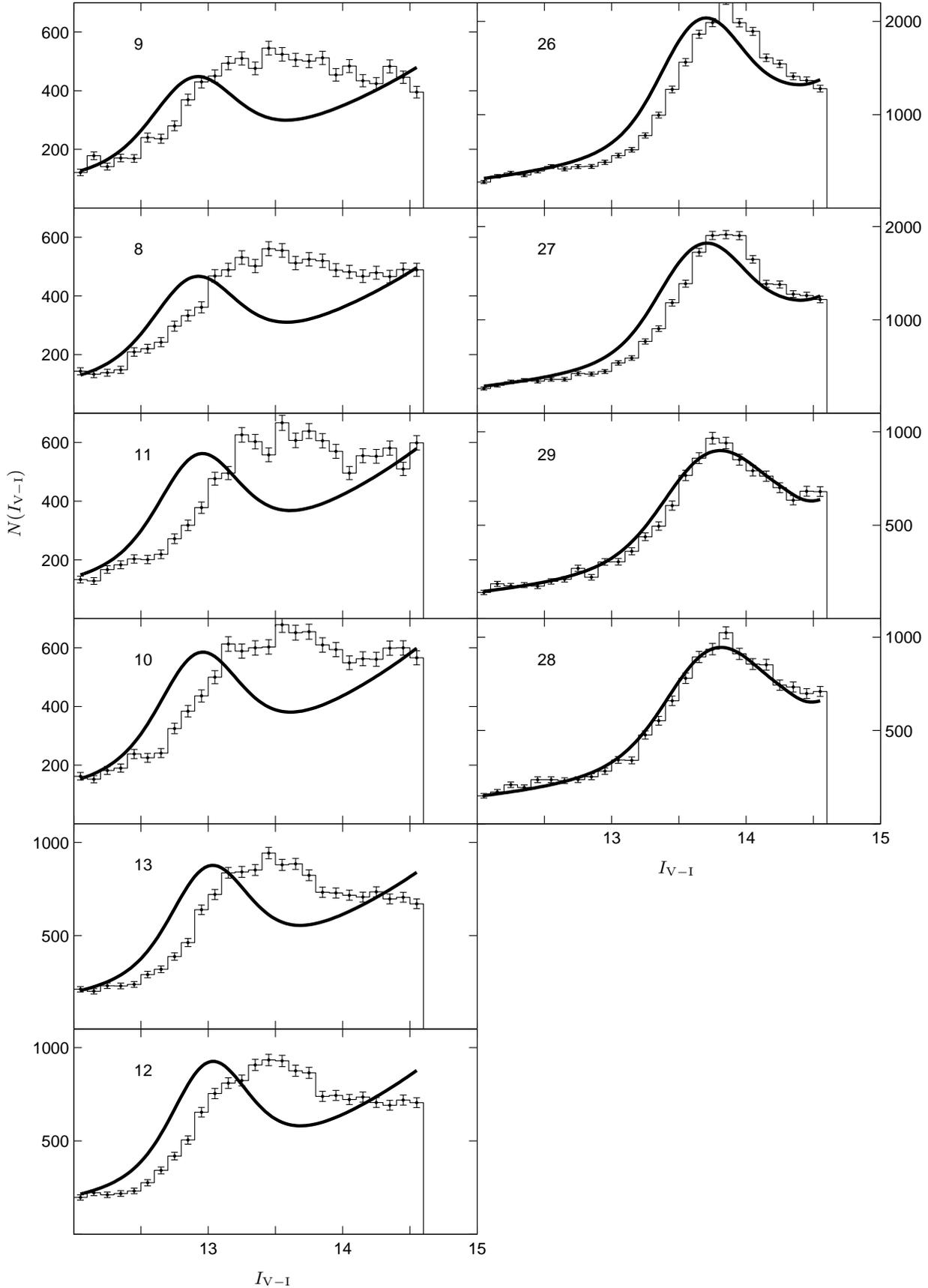}}
\caption{\label{fig:histMatrixWide}Predicted number count profiles using the best-fitting E3 model (equation~(\ref{eq:numcount})) with the observed number count histograms for OGLE-II fields with $l> 6^{\circ}$ (left column) and $l< -4^{\circ}$ (right column). The predicted number count profiles are significantly different for fields with  $l> 6^{\circ}$, however the predicted magnitude peak of the red clump roughly coincides with those observed for fields with $l< -4^{\circ}$.}
\end{figure*}

The OGLE-II data from fields with longitude $l< -4^{\circ}$ and $l> 6^{\circ}$ were excluded from the original bar modelling based on the findings illustrated by Fig.~\ref{fig:distmod}, and because initial modelling efforts using all data from all fields simultaneously failed to produce satisfactory results. Using the best-fitting model (E3) determined using the central OGLE-II fields, Fig.~\ref{fig:histMatrixWide} shows the predicted number count densities for fields with $l< -4^{\circ}$ and $l> 6^{\circ}$, overlaid on the observed number count histograms.

\begin{figure}
\psfrag{xlabel}{\normalsize $x$ $(\kpc)$}
\psfrag{ylabel}{\normalsize $y$ $(\kpc)$}
\includegraphics[width=1.0\hsize]{\FigDir{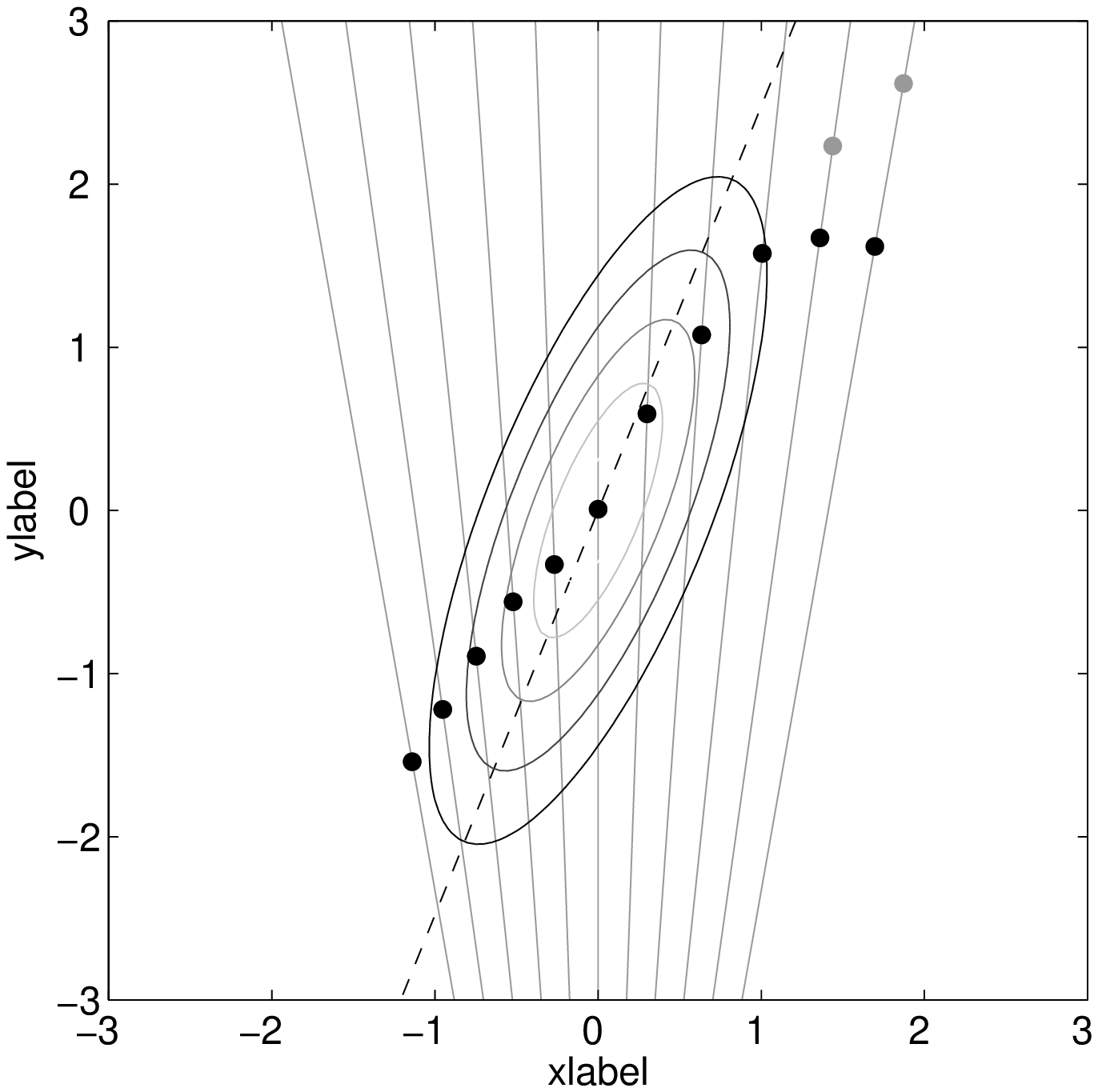}}
\caption{\label{fig:midplaneScan}Mean position of the red clump determined using the best-fitting analytic bar model (E3). Gray lines indicate 11 lines-of-sight through the E3 model, with $l=-10^{\circ}, -9.8^{\circ},\ldots,10^{\circ}$ and $b=0.0^{\circ}$ and contours indicating the density profile of the model. The dashed line shows the orientation of the bar major axis, with the Sun positioned at $(0,-8)$ \kpc. Number count profiles are generated using equation~(\ref{eq:numcount}). The mean red clump magnitude for each line-of-sight was converted to a distance, and plotted as solid circles. Black circles indicate the mean position of the red clump when the exponential cut-off, equation~(\ref{eq:cutoff}) is applied. Grey circles indicate the mean RCG position when this cut-off is not imposed. The observed features in Fig.~\ref{fig:distmod} at longitudes $|l| \gtrsim 5^{\circ}$ are more clearly reproduced when the cut-off is applied to the analytic model.}
\end{figure}

It is clear from the left hand column of Fig.~\ref{fig:histMatrixWide} that the mean position of the red clump in fields with $l> 6^{\circ}$ are significantly removed from that predicted by the E3 linear bar model. However, the right hand column of Fig.~\ref{fig:histMatrixWide} shows that the observed position of the red clump is in rough agreement with that predicted from the model for fields with $l< -4^{\circ}$. We test this further, by taking lines of sight through the best-fitting E3 model and computing the number count profile using equation~(\ref{eq:numcount}). The mean magnitude of the red clump is converted to a distance for each line of sight. Fig.~\ref{fig:midplaneScan} shows the density contours of the analytic E3 model overlaid with the mean position of the red clump determined in this way.
The features of Fig.~\ref{fig:distmod} at longitudes $|l| \gtrsim 5^{\circ}$ are qualitatively reproduced in Fig.~\ref{fig:midplaneScan}. This suggests that the observed data are likely to be consistent with a single bar-like structure, rather than requiring additional structure elements. We note however, that the abrupt departure of the location of the density peak in  Fig.~\ref{fig:distmod} at longitudes $l \lesssim 5^{\circ}$ is best reproduced in Fig.~\ref{fig:midplaneScan} by the analytical  density model when the exponential cut-off of equation~(\ref{eq:cutoff}) is imposed. 

Fields with $l< -4^{\circ}$ and $l> 6^{\circ}$ should be included in the modelling of the bar, as they are likely to provide further constraints on the final model. Initial attempts to model the bar using all fields failed because the fitting algorithm found a model with a bar angle of $45^{\circ}$, consistent with the observed data, but not consistent with the current understanding of the bar. Fig.~\ref{fig:midplaneScan} shows that the morphology indicated by the observed data in Fig.~\ref{fig:distmod} can be explained using a bar structure oriented at $\sim 20^{\circ}$ to the Sun-GC line-of-sight.  \citet{2007astro.ph..2109C} also note that the line-of-sight density for tri-axial bulges reaches a maximum where the line-of-sight is tangential to the ellipsoidal density contours. These authors also quantify the difference between the positions of maximum density and the intersection of the line-of-sight with the major axis of the bar.

\begin{figure*}
\psfrag{xlabel}{\normalsize \raisebox{-5pt} \IVmI}
\psfrag{ylabel}{\normalsize \raisebox{5pt} {$N(\IVmI)$}}
\includegraphics[height=0.95\textheight]{\FigDir{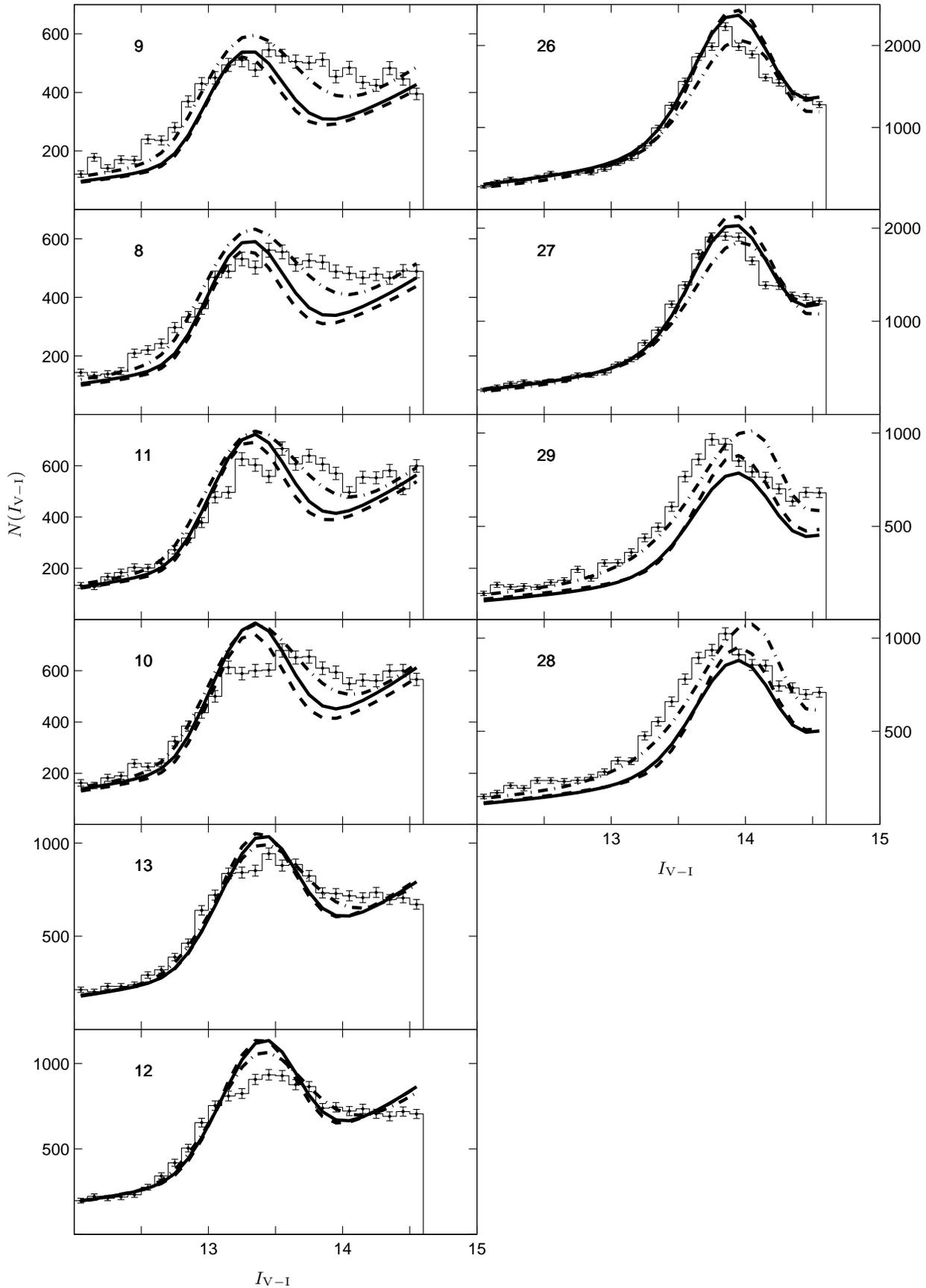}}
\caption{\label{fig:histMatrixWide_G}Best-fitting number count profiles using the Gaussian `G' type tri-axial models  with the observed number count histograms for OGLE-II fields with $l> 6^{\circ}$ (left column) and $l< -4^{\circ}$ (right column). Solid, dashed and dot-dashed lines correspond to model subtype 1, 2 and 3 respectively. }
\end{figure*}

The fitting procedure for all nine tri-axial models was repeated, including now the 10 OGLE-II fields with $l< -4^{\circ}$ or $l> 6^{\circ}$.   Figure~\ref{fig:histMatrixWide_G} shows the best-fitting number count profiles for the ten non-central fields for the `G' type of analytic tri-axial model\footnote{Similar figures showing the best-fitting models using the `E' and `P' type density profiles appear in the on-line supplementary material.}. The best-fitting number count profiles for fields with $-4^{\circ} \leq l \leq 6^{\circ}$ are not significantly different to those shown in Figure~\ref{fig:matrixE2}. The main features of the observed number count profiles are reproduced by the best-fitting analytic models. There are however instances where details of the observed number counts are not reproduced. Magnitude bins with $\IVmI \gtrsim 13$ are  poorly fitted by the G-type models in positive longitude fields. Two of the E-type models (E2, E3) similarly fail to trace these data. The number count profile for model E1 shows a flattened peak, resulting from the pronounced box-like nature of the density profile. All the P-type models appear to reproduce the data at these magnitudes with similar profiles. All models predict a RCG peak at magnitudes greater than that observed in fields 28 and 29. This offset between predicted and observed RCG peak location is also seen in fields 26 and 27 but to a lesser degree.

The best-fitting parameters are shown in Table~\ref{tab:paramValsWide}. The model which results in the lowest value of \chisq is still E3. Model G2 also remains the worst-fitting model to the data. 

\begin{table*}
\caption{\label{tab:paramValsWide}Best fitting parameter values for all density models (equations~\ref{eq:G1}-\ref{eq:P3}) fitted to the number count histograms from all 44 OGLE-II fields.}
\begin{tabular}{cccccccc}
\hline
& \multicolumn{2}{c}{Bar orientation $(^{\circ})$} & \multicolumn{3}{c}{Scale lengths (\pc)} & & Axis ratios  \\ 
Model & $\beta$ &  $\alpha'$ & $x_{0}$ & $y_{0}$ & $z_{0}$   & \chisq & $x_{0}:y_{0}:z_{0}$ \\
\hline
\input{paramTexTableWide.txt}
\hline
\end{tabular}
\end{table*}

\begin{figure}
\psfrag{xlabel}{\raisebox{-5pt} \normalsize Model}
\psfrag{ylabel}{\normalsize Ratio}
\psfrag{XXXXXXXXXXXXX}{\small $x_{0,M=44} / x_{0,M=34}$}
\psfrag{second}{\small $y_{0,M=44} / y_{0,M=34}$}
\psfrag{third}{\small $z_{0,M=44}/z_{0,M=34}$}
\includegraphics[width=1.0\hsize]{\FigDir{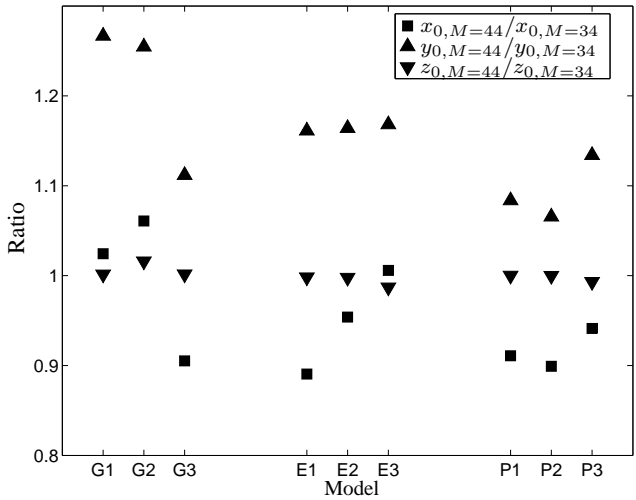}}
\caption{\label{fig:paramChange}Ratio of bar scale lengths from best-fitting models using data from all 44 OGLE-II fields to those using data from the central 34 OGLE-II fields.}
\end{figure}

We now consider the change to the best-fitting model parameters when data from all fields are used in the analysis. The position angle of the bar semi-major axis with respect to the Sun-Galactic centre line-of-sight is now $\alpha' = 24^{\circ}$ -- $27^{\circ}$. The bar scale lengths $x_{0}$, $y_{0}$ and $z_{0}$ also changed; Fig.~\ref{fig:paramChange} shows the ratio of the best-fitting scale lengths determined using all 44 fields to those found using only the central 34 fields. On average, upon including the data from wide longitude fields, the semi-major axis scale length, $x_{0}$, decreased by $\sim 5\%$; the semi-minor axis scale length, $y_{0}$, increased by $\sim 16\%$ and the vertical scale length, $z_{0}$, remained essentially unchanged. The relatively large change in the semi-minor axis scale length is intuitively understandable, as a bar position angle of $\sim 25\deg$ with respect to the Sun-GC line-of-sight mean the semi-minor axis direction has a large vector component in the Galactic longitude direction. The additional constraints of the data from fields at extended Galactic longitudes can therefore have a pronounced effect on the fitted values of $y_{0}$. Similarly, we expect some weak correlation between the semi-major and semi-minor axis scale lengths, evidenced by the slight decrease on average for the $x_{0}$ values upon adding the data from wide longitude fields. It is unsurprising that the values of $z_{0}$ are unchanged, as adding data from the wide longitude fields has little power in further constraining model elements in the direction perpendicular to the Galactic midplane.

As above, we consider the ratio of the axis scale lengths for each model (see Fig.~\ref{fig:axisRatios}). The ratio $x_{0}/y_{0}$ for all nine models using all field data has values 2.7 -- 3.0. This range is now completely consistent with the range 2.5 -- 3.3 reported by \citet{2002MNRAS.330..591B}. The ratio $x_{0}/z_{0}$ now lies in the range 3.6 -- 3.9, with the exception of that for model G1 which has $x_{0}/z_{0} = 5.9$. The range $x_{0}/z_{0} = $ 3.6 -- 3.9 is narrower than that found previously, 3.4 -- 4.2, yet still higher on average than 2.8 -- 3.8 reported by \citet{1997ApJ...477..163S}. The scale length ratios are now $x_{0}:y_{0}:z_{0} = 10:3.5:2.6$, which compared to those determined using only the central OGLE-II fields ($x_{0}:y_{0}:z_{0} = 10:3.0:2.6$) are closer to the working model proposed by \citet{2002ASPC..273...73G} which has $x_{0}:y_{0}:z_{0} = 10:4:3$.

There are features in the number count histograms that are not faithfully reproduced by the analytic tri-axial bar models for wide longitude fields. Specifically, the predicted number count dispersions around the RCG peak magnitudes for fields with $l>6\deg$ are too small compared to that observed, see Fig.~\ref{fig:histMatrixWide_G}. While the location of the observed maximum line-of-sight density can be reproduced using just a bar, it is possible that the reason why the analytic bar models underestimate the observed number count dispersions for fields with $l>6\deg$ is because these models do not include elements which correspond to extended aggregations of stars at or near the ends of the bar. Clumps of stars at the ends of the bar would increase the line-of-sight density dispersion at wide longitudes. Such extra aggregations of stars have been observed in galaxies with boxy or peanut-shaped bulges \citep{2006MNRAS.370..753B}. Their origin may be due to the superposition of members of the $x_{1}$ family of orbits \citep{2002MNRAS.337..578P,1987gady.book.....B} many of which have loops near the end of the bar. Alternatively, the aggregations of stars near the ends of the bar may be the edge-on projection of an inner ring (\citealt{2006MNRAS.370..753B} and references therein). The presence of another density structure such as a long thin bar oriented at $\sim 45\deg$ \citep{2007astro.ph..2109C} might also contribute to the relatively large line-of-sight density dispersion. The effects of such a structure would be most pronounced at wide longitudes. The presence of a spiral arm might similarly contribute to the large line-of-sight density dispersion for these fields. 

\subsection{Varying $R_{0}$}
\label{sec:changeR0}
We consider in this section the effect on the fitted model parameters of changing the Galactocentric distance $R_{0}$. It was noted in Section~\ref{sec:popeffects} that there is an offset between the peak magnitude of RCGs observed in the OGLE-II data and that expected for a RCG, with absolute magnitude similar to that of local RCGs, placed at a distance of 8.0 \kpc. The observed RCGs are systematically 0.3 mag fainter than the fiducial local RCG at 8.0 \kpc. There are two possible reasons for this magnitude offset. Firstly, the adopted Galactocentric distance of $R_{0} = 8.0$ \kpc may be incorrect; secondly, there may be population variations between local and bulge RCGs, resulting in different intrinsic magnitudes in the two populations. In Section~\ref{sec:popeffects} we noted that the offset predicted from the theoretical population models of  \citet{2001MNRAS.323..109G} estimate $\DelIRC \simeq -0.1$. We postulated that if the remaining magnitude offset is completely accounted for by a change in the Galactocentric distance, this would mean that $R_{0} = 7.3$ \kpc. Conversely, if we adopted the value of $R_{0} = 7.6\pm0.32$ \kpc determined by \citet{2005ApJ...628..246E}, the magnitude offset would become $-0.18$ mag, where including the population effect predicted by \citet{2001MNRAS.323..109G} would leave an unaccounted-for offset of $-0.08$ mag. In this section we present the results of further modelling where we apply magnitude offsets which correspond to different values of $R_{0}$.

In Section~\ref{sec:results}, we note that for the OGLE-II fields 6, 7, 14, 15 and 43, which have latitudes removed from the majority of OGLE-II fields, there are systematic offsets between the RCG peak in observed number count histograms and the predictions of all nine tri-axial models tested. For this reason, we limited the fields used to those which have Galactic latitude $-5\deg \leq b \leq -2\deg$. We also noted in Section~\ref{sec:results} that due to the lack of field coverage in the latitude direction the data are not effective at constraining the bar tilt angle $\beta$. For this reason, $\beta$ was held constant at $0.0\deg$ in the following modelling analysis. During the previous modelling efforts of  Section~\ref{sec:results} it was also found that applying the exponential cut-off of equation~(\ref{eq:cutoff}) had little effect on determining the best-fitting bar parameters. This cut-off was not applied in the following analysis.

The best-fitting bar parameters were determined for the nine tri-axial bar models for the OGLE-II fields occupying the latitude strip $-5\deg \leq b \leq -2\deg$ using magnitude offsets corresponding to $R_{0}$ values of 8.0 \kpc, 7.6 \kpc and 7.3 \kpc. The resulting best-fitting parameters are listed in Table~\ref{tab:changeR0}.

\begin{table*}
\caption{\label{tab:changeR0}Best-fitting parameters for all density models (equations~\ref{eq:G1}-\ref{eq:P3}) fitted to the number count histograms for OGLE-II fields with Galactic latitude $-5\deg \leq b \leq -2\deg$. Three values of Galactocentric distance $R_{0}$ were used. The bar tilt angle, $\beta$, was held at 0.0\deg.}
\begin{tabular}{cccccccc}
\hline
\multicolumn{3}{c}{} & \multicolumn{3}{c}{Scale lengths (\pc)} & & Axis ratios \\
Model & $R_{0}$ (\kpc) & $\alpha$ (\deg) & $x_{0}$ & $y_{0}$ & $z_{0}$ & \chisq & $x_{0}:y_{0}:z_{0}$ \\
\hline
\input{paramTexTableChangeR0.txt}
\hline
\end{tabular}
\end{table*}

From the results in Table~\ref{tab:changeR0} we see that the fitted bar opening angle $\alpha$ increases for all models as $R_{0}$ decreases. The fitted scale lengths $x_{0}$, $y_{0}$ and $z_{0}$ all decrease linearly as $R_{0}$ decreases. The values of \chisq in Table~\ref{tab:changeR0} indicate that a smaller value of $R_{0}$ is favoured by all models. The ratio between the scale heights remains remarkably constant with varying $R_{0}$, indicating that while the orientation of the bar changes slightly with varying $R_{0}$, the overall shape of the bar does not change. 

The mean values of the scale length ratios for all models, and all values of $R_{0}$ are $x_{0}:y_{0}:z_{0} = 10 : 3.6 : 2.7$. This result is very close to that determined using data from all 44 OGLE-II fields, see Section~\ref{sec:results} above. 

\section{Discussion}
\label{sec:discussion}
Red clump giant stars in the OGLE-II microlensing survey catalogue can be used as tracers of the bulge density over a large region towards the Galactic centre. Nine analytic tri-axial bar models were initially fitted to the number count histograms of red clump stars observed in 34 OGLE-II fields which have $-4^{\circ} \leq l \leq 6^{\circ}$. The models all have the major axis of the bar oriented at $20^{\circ}$ -- $26^{\circ}$ to the Sun-Galactic centre line-of-sight. This orientation is in agreement with the results of \citet{1997ApJ...477..163S} and \citet{1997ApJ...487..885N} which give a bar angle of $20^{\circ}$ -- $30^{\circ}$, and is marginally in agreement with the value of $12\pm6^{\circ}$ from \citet{2000MNRAS.313..392L}. \citet{2002MNRAS.330..591B} obtain best-fitting non-parametric models to the COBE/DIRBE L band map of the inner Galaxy with bar angles of $20^{\circ}$ -- $25^{\circ}$.

We find the ratio of the bar major axis scale length to minor axis scale length in the plane of the Galaxy to be $x_{0}/y_{0} = 3.2$ -- $3.6$, higher than the value of 2.0 -- 2.4 reported by \citet{1997ApJ...477..163S}, but consistent with the upper end of the range 2.5 -- 3.3 found by \citet{2002MNRAS.330..591B}. The ratio of bar major axis scale length to vertical axis scale length was found to be $x_{0}/z_{0}=3.4$ -- 4.2, again higher on average than that reported by \citet{1997ApJ...477..163S} who found $x_{0}/z_{0}=2.8$ -- 3.8, and higher than the value of $\simeq 3.3$ generally adopted \citep{2002ASPC..273...73G}. The working model proposed by \citet{2002ASPC..273...73G} gives the scale length ratios as  $x_{0}:y_{0}:z_{0} = 10:4:3$. Our results suggest a more prolate model with  $x_{0}:y_{0}:z_{0} = 10:3.0:2.6$. 

The observed separation of the mean position of red clump giants from the bar major axis at $|l| \gtrsim 5^{\circ}$ was shown to be a geometric effect, rather than evidence of a more complicated structure such as a ring. The observed data from these fields were used to further constrain the bar models. The resulting bar position angles was found to be $24^{\circ}$ -- $27^{\circ}$. This narrower range is consistent with the several values of the bar position angle found by previous studies. The bar scale length ratios were determined to be $x_{0}:y_{0}:z_{0} = 10:3.5:2.6$, which are closer to the working model proposed by \citet{2002ASPC..273...73G} than those values found using data from only the central 34 OGLE-II fields.

Reasons for the difference between the bar axis ratios determined here and the general working model of \citet{2002ASPC..273...73G} may include RCG population effects noted above in Section~\ref{sec:popeffects}. The intrinsic luminosity of bulge RCG stars was assumed to be the same as the local population, however it was found that an offset of $-0.3$ mag had to be applied to the computed distance moduli in order to obtain results consistent with standard bar models. \citet{2004MNRAS.349..193S} found the mean magnitude of observed bulge clump stars is found to be 0.3 mag fainter to that expected assuming (i) the properties of bulge RCGs are the same as the local population and (ii) the distance to the Galactic centre is 8 \kpc. A possible implication is that the adopted distance to the Galactic centre of 8 \kpc may be incorrect. The bar modelling procedure was repeated for three values of the Sun-Galactic centre distance $R_{0}$: 7.3 \kpc, 7.6 \kpc and 8.0 \kpc, using data from OGLE-II fields which have $-5\deg \leq b \leq -2\deg$. Some fields with latitudes outside this range were found to have number count histograms which were not reproducible by any linear tri-axial model of the bar tested in this work, and were therefore excluded. In addition, most of the OGLE-II fields latitudes $-4^{\circ} < b < -2^{\circ}$ were excluded. The low amount of information in the latitude direction means that the data have little leverage in determining some of the model parameters.  Without more data in these regions, model fitting algorithms may not be able to refine some model parameters such as the tilt angle $\beta$. The three values of $R_{0}$ used are the `default' value of 8.0 \kpc; 7.6 \kpc, as determined by \citet{2005ApJ...628..246E}; and 7.3 \kpc, which corresponds to the value of $R_{0}$ consistent with the observed mean RCG magnitudes in the OGLE-II bulge fields assuming the population effects of \citet{2001MNRAS.323..109G}. The fitted scale lengths $x_{0}$, $y_{0}$ and $z_{0}$ were found to increase linearly with increasing values of $R_{0}$, while the bar opening angle $\alpha$ decreased slightly with increasing $R_{0}$. The shape of the bar, as quantified by the ratio of axis scale lengths, was found to be insensitive to different values of $R_{0}$. The mean ratio of scale lengths over all models and values of $R_{0}$ was found to be $x_{0}:y_{0}:z_{0} = 10:3.6:2.7$, which is slightly closer to the working model of \citet{2002ASPC..273...73G} for the Galactic bar than that determined using all 44 OGLE-II fields. The goodness-of-fit measure \chisq decreased for all models as the value of  $R_{0}$ was lowered.   

Improved modelling for the Galactic bar may be possible through the addition of further elements to the methods presented here. The possibility of a metallicity gradient across the bulge was not accounted for in this work. Including spiral terms (see e.g. \citealt{2002ApJ...567L.119E}) in the analytic density profiles may similarly result in a closer reproduction of the observed number count profiles. 
 
The observed maximum line-of-sight density can be reproduced using just a bar, without requiring additional structure elements. However, the finer details of the number count histograms, especially at wide longitudes, were not reproduced by the tri-axial bar models used here. The predicted number count dispersions around the peak RCG magnitude were underestimated by the analytical models, particularly for fields with $l>6\deg$. It is possible that these finer features can be reproduced using models which include extra stellar agglomerations at the ends of the bar. The additional stellar densities could arise from specific stellar orbits aligned with the bar, or due to the projection effect of an inner ring. A long thin bar as postulated by \citet{2007astro.ph..2109C} might similarly increase the line-of-sight density dispersion, particularly for wide longitude fields, as might the presence of a spiral arm. Modelling the bar using non-parametric methods (see e.g. \citealt{1988MNRAS.232..431E}) may provide valuable insight into the existence and nature of such additional features. Some preliminary work applying these methods has begun.

Data from current and future surveys of the Galactic bulge region will be useful for refining the constraints on the bar parameters. The third evolution of the OGLE microlensing experiment, OGLE-III, is currently in progress, covering a larger region of the central Galactic region than OGLE-II. 

Infra-red observations of the bulge have the advantage of lower extinction effects due to dust, compared to optical observations. Current catalogues which would be suitable for investigating the structure of the Galactic bar include the point source catalogue from the 2MASS All Sky data release \citep{2006AJ....131.1163S}, the Galactic Plane Survey from the UKIRT Infra-Red Deep Sky Survey \citep{Law06,2006MNRAS.372.1227D} and data from the ISOGAL \citep{2003A&A...403..975O}, Spitzer/GLIMPSE \citep{2005ApJ...630L.149B} and AKARI \citep{2006IAUJD..13E..11I} space telescope surveys. Proposed infra-red surveys towards the Galactic centre include the Galactic Bar Infra-red Time-domain (GABARIT) survey which aims to monitor the Galactic bar region in the K band for microlensing events (Kerins 2006, private communication). The analysis of red clump giant star counts in these surveys may result in tighter constraints on the properties of the bar, and can be combined with constrains on stellar kinematics from proper motion surveys \citep{Rat06} in order to develop dynamical models of the inner Galaxy.

\section*{Acknowledgements}

We thank D. Faria, {\L}. Wyrzykowski, R. James and W. Evans for helpful discussions. NJR acknowledges financial support by a PPARC PDRA fellowship and the LKBF. NJR thanks the Kapteyn Astronomical Institute, RuG, for visitor support.  This work was partially supported by the European Community's Sixth Framework Marie Curie Research Training Network Programme, Contract No. MRTN-CT-2004-505183 `ANGLES'.


\end{document}

%% file: DisModtexTable.txt
  1 &   1.08 &  -3.62 &   13.616 \ppm 0.005 &   0.2936 \ppm  0.0043 & 0.964 \ppm  0.02 &    14.55 \ppm  0.09  & 0.21 &  31002 \\ 
  2 &   2.23 &  -3.46 &   13.536 \ppm 0.005 &   0.3130 \ppm  0.0040 & 0.964 \ppm  0.02 &    14.47 \ppm  0.09  & 0.24 &  33813 \\ 
  3 &   0.11 &  -1.93 &   13.664 \ppm 0.003 &   0.2461 \ppm  0.0026 & 0.964 \ppm  0.04 &    14.60 \ppm  0.09  & 0.14 &  66123 \\ 
  4 &   0.43 &  -2.01 &   13.655 \ppm 0.003 &   0.2517 \ppm  0.0026 & 0.964 \ppm  0.04 &    14.59 \ppm  0.09  & 0.15 &  65748 \\ 
  5 &  -0.23 &  -1.33 &   13.630 \ppm 0.003 &   0.2809 \ppm  0.0030 & 0.964 \ppm  0.06 &    14.56 \ppm  0.10  & 0.20 &  43493 \\ 
  6 &  -0.25 &  -5.70 &   13.606 \ppm 0.011 &   0.4197 \ppm  0.0088 & 0.964 \ppm  0.03 &    14.54 \ppm  0.09  & 0.37 &  12085 \\ 
  7 &  -0.14 &  -5.91 &   13.589 \ppm 0.012 &   0.4255 \ppm  0.0100 & 0.964 \ppm  0.03 &    14.52 \ppm  0.09  & 0.38 &  11328 \\ 
  8 &  10.48 &  -3.78 &   13.366 \ppm 0.002 &   1.4277 \ppm  0.0013 & 0.964 \ppm  0.03 &    14.30 \ppm  0.09  & 1.41 &  10248 \\ 
  9 &  10.59 &  -3.98 &   13.383 \ppm 0.012 &   0.5855 \ppm  0.0110 & 0.964 \ppm  0.03 &    14.32 \ppm  0.09  & 0.55 &   9971 \\ 
 10 &   9.64 &  -3.44 &   13.407 \ppm 0.002 &   1.4178 \ppm  0.0013 & 0.964 \ppm  0.03 &    14.34 \ppm  0.09  & 1.40 &  12068 \\ 
 11 &   9.74 &  -3.64 &   13.438 \ppm 0.015 &   0.3902 \ppm  0.0125 & 0.964 \ppm  0.04 &    14.37 \ppm  0.09  & 0.33 &  11345 \\ 
 12 &   7.80 &  -3.37 &   13.381 \ppm 0.008 &   0.4692 \ppm  0.0072 & 0.964 \ppm  0.04 &    14.31 \ppm  0.09  & 0.42 &  15936 \\ 
 13 &   7.91 &  -3.58 &   13.389 \ppm 0.009 &   0.4234 \ppm  0.0081 & 0.964 \ppm  0.03 &    14.32 \ppm  0.09  & 0.37 &  15698 \\ 
 14 &   5.23 &   2.81 &   13.550 \ppm 0.006 &   0.3131 \ppm  0.0053 & 0.964 \ppm  0.04 &    14.48 \ppm  0.09  & 0.24 &  27822 \\ 
 15 &   5.38 &   2.63 &   13.564 \ppm 0.007 &   0.3027 \ppm  0.0059 & 0.964 \ppm  0.04 &    14.50 \ppm  0.09  & 0.23 &  24473 \\ 
 16 &   5.10 &  -3.29 &   13.487 \ppm 0.007 &   0.3200 \ppm  0.0057 & 0.964 \ppm  0.03 &    14.42 \ppm  0.09  & 0.25 &  22055 \\ 
 17 &   5.28 &  -3.45 &   13.474 \ppm 0.007 &   0.3270 \ppm  0.0058 & 0.964 \ppm  0.03 &    14.41 \ppm  0.09  & 0.26 &  23132 \\ 
 18 &   3.97 &  -3.14 &   13.471 \ppm 0.005 &   0.2983 \ppm  0.0044 & 0.964 \ppm  0.02 &    14.40 \ppm  0.09  & 0.22 &  32457 \\ 
 19 &   4.08 &  -3.35 &   13.491 \ppm 0.006 &   0.3026 \ppm  0.0048 & 0.964 \ppm  0.03 &    14.42 \ppm  0.09  & 0.23 &  30410 \\ 
 20 &   1.68 &  -2.47 &   13.583 \ppm 0.004 &   0.2726 \ppm  0.0031 & 0.964 \ppm  0.03 &    14.52 \ppm  0.09  & 0.18 &  49900 \\ 
 21 &   1.80 &  -2.66 &   13.596 \ppm 0.004 &   0.2886 \ppm  0.0034 & 0.964 \ppm  0.02 &    14.53 \ppm  0.09  & 0.21 &  45578 \\ 
 22 &  -0.26 &  -2.95 &   13.741 \ppm 0.004 &   0.2669 \ppm  0.0034 & 0.964 \ppm  0.04 &    14.67 \ppm  0.09  & 0.18 &  42914 \\ 
 23 &  -0.50 &  -3.36 &   13.724 \ppm 0.004 &   0.2778 \ppm  0.0037 & 0.964 \ppm  0.04 &    14.66 \ppm  0.09  & 0.19 &  36030 \\ 
 24 &  -2.44 &  -3.36 &   13.817 \ppm 0.004 &   0.2638 \ppm  0.0037 & 0.964 \ppm  0.04 &    14.75 \ppm  0.09  & 0.17 &  35351 \\ 
 25 &  -2.32 &  -3.56 &   13.810 \ppm 0.004 &   0.2695 \ppm  0.0039 & 0.964 \ppm  0.03 &    14.74 \ppm  0.09  & 0.18 &  31801 \\ 
 26 &  -4.90 &  -3.37 &   13.815 \ppm 0.005 &   0.2897 \ppm  0.0046 & 0.964 \ppm  0.02 &    14.75 \ppm  0.09  & 0.21 &  26940 \\ 
 27 &  -4.92 &  -3.65 &   13.794 \ppm 0.006 &   0.2782 \ppm  0.0050 & 0.964 \ppm  0.02 &    14.73 \ppm  0.09  & 0.19 &  24603 \\ 
 28 &  -6.76 &  -4.42 &   13.785 \ppm 0.010 &   0.3081 \ppm  0.0082 & 0.964 \ppm  0.02 &    14.72 \ppm  0.09  & 0.23 &  13702 \\ 
 29 &  -6.64 &  -4.62 &   13.762 \ppm 0.009 &   0.2792 \ppm  0.0083 & 0.964 \ppm  0.02 &    14.70 \ppm  0.09  & 0.19 &  12893 \\ 
 30 &   1.94 &  -2.84 &   13.570 \ppm 0.004 &   0.2746 \ppm  0.0034 & 0.964 \ppm  0.03 &    14.50 \ppm  0.09  & 0.19 &  41748 \\ 
 31 &   2.23 &  -2.94 &   13.535 \ppm 0.004 &   0.2892 \ppm  0.0036 & 0.964 \ppm  0.02 &    14.47 \ppm  0.09  & 0.21 &  40623 \\ 
 32 &   2.34 &  -3.14 &   13.528 \ppm 0.004 &   0.3001 \ppm  0.0038 & 0.964 \ppm  0.02 &    14.46 \ppm  0.09  & 0.22 &  35954 \\ 
 33 &   2.35 &  -3.66 &   13.559 \ppm 0.005 &   0.3206 \ppm  0.0045 & 0.964 \ppm  0.02 &    14.49 \ppm  0.09  & 0.25 &  30882 \\ 
 34 &   1.35 &  -2.40 &   13.608 \ppm 0.003 &   0.2711 \ppm  0.0031 & 0.964 \ppm  0.03 &    14.54 \ppm  0.09  & 0.18 &  52216 \\ 
 35 &   3.05 &  -3.00 &   13.533 \ppm 0.005 &   0.3049 \ppm  0.0040 & 0.964 \ppm  0.02 &    14.47 \ppm  0.09  & 0.23 &  36796 \\ 
 36 &   3.16 &  -3.20 &   13.508 \ppm 0.005 &   0.3104 \ppm  0.0042 & 0.964 \ppm  0.02 &    14.44 \ppm  0.09  & 0.24 &  34437 \\ 
 37 &   0.00 &  -1.74 &   13.636 \ppm 0.003 &   0.2488 \ppm  0.0025 & 0.964 \ppm  0.05 &    14.57 \ppm  0.10  & 0.15 &  72098 \\ 
 38 &   0.97 &  -3.42 &   13.637 \ppm 0.005 &   0.2911 \ppm  0.0038 & 0.964 \ppm  0.02 &    14.57 \ppm  0.09  & 0.21 &  34675 \\ 
 39 &   0.53 &  -2.21 &   13.687 \ppm 0.003 &   0.2524 \ppm  0.0027 & 0.964 \ppm  0.04 &    14.62 \ppm  0.09  & 0.15 &  60217 \\ 
 40 &  -2.99 &  -3.14 &   13.854 \ppm 0.004 &   0.2459 \ppm  0.0036 & 0.964 \ppm  0.04 &    14.79 \ppm  0.09  & 0.14 &  35426 \\ 
 41 &  -2.78 &  -3.27 &   13.857 \ppm 0.004 &   0.2543 \ppm  0.0035 & 0.964 \ppm  0.04 &    14.79 \ppm  0.09  & 0.16 &  34118 \\ 
 42 &   4.48 &  -3.38 &   13.494 \ppm 0.006 &   0.3354 \ppm  0.0051 & 0.964 \ppm  0.03 &    14.43 \ppm  0.09  & 0.27 &  27377 \\ 
 43 &   0.37 &   2.95 &   13.839 \ppm 0.004 &   0.2654 \ppm  0.0033 & 0.964 \ppm  0.05 &    14.77 \ppm  0.10  & 0.17 &  40730 \\ 
 45 &   0.98 &  -3.94 &   13.595 \ppm 0.006 &   0.3302 \ppm  0.0047 & 0.964 \ppm  0.03 &    14.53 \ppm  0.09  & 0.26 &  29009 \\ 
 46 &   1.09 &  -4.14 &   13.616 \ppm 0.006 &   0.3189 \ppm  0.0047 & 0.964 \ppm  0.03 &    14.55 \ppm  0.09  & 0.25 &  26027 \\ 

%% file: paramTexTable.txt
G1 &  0.00 & 21.82 & 1469.00 &  449.28 &  391.80 & 15302.31 & \\
   & -0.99 & 21.82 & 1469.41 &  448.78 &  391.62 & 15282.57 & 10.0 : 3.1 : 2.7 \\
G2 &  0.00 & 19.76 & 1206.92 &  375.94 &  353.81 & 16558.28 & \\
   & -0.28 & 19.79 & 1203.37 &  377.22 &  354.24 & 16547.35 & 10.0 : 3.1 : 2.9 \\
G3 &  0.00 & 25.57 & 5289.91 & 1512.32 & 1277.52 & 14306.20 & \\
   & -0.30 & 25.55 & 5288.64 & 1511.72 & 1277.98 & 14306.06 & 10.0 : 2.9 : 2.4 \\
E1 &  0.00 & 21.63 & 2143.51 &  540.08 &  325.49 & 15766.77 & \\
   &  0.06 & 21.62 & 2135.11 &  539.86 &  325.48 & 15753.98 & 10.0 : 2.5 : 1.5 \\
E2 &  0.00 & 23.68 & 1034.30 &  306.39 &  261.43 & 11930.61 & \\
   & -0.01 & 23.68 & 1034.24 &  306.38 &  261.47 & 11930.58 & 10.0 : 3.0 : 2.5 \\
E3 &  0.00 & 21.92 & 1039.86 &  323.80 &  299.08 & 10722.69 & \\
   &  0.47 & 21.97 & 1039.62 &  323.75 &  298.75 & 10711.13 & 10.0 : 3.1 : 2.9 \\
P1 &  0.00 & 24.66 & 1988.33 &  562.49 &  478.93 & 13952.47 & \\
   &  0.08 & 24.66 & 1988.10 &  562.39 &  478.84 & 13952.09 & 10.0 : 2.8 : 2.4 \\
P2 &  0.00 & 25.07 & 3906.98 & 1088.90 &  927.00 & 15183.47 & \\
   &  0.76 & 25.03 & 3907.42 & 1089.36 &  927.61 & 15180.89 & 10.0 : 2.8 : 2.4 \\
P3 &  0.00 & 23.81 & 1992.91 &  580.32 &  491.30 & 12123.07 & \\
   &  0.10 & 23.82 & 1993.85 &  580.33 &  491.09 & 12122.82 & 10.0 : 2.9 : 2.5 \\

%% file: paramTexTableWide.txt
G1 &  0.75 & 27.06 & 1505.20 &  568.49 &  392.19 & 23808.97 & 10.0 : 3.8 : 2.6 \\
G2 & -7.68 & 24.49 & 1276.56 &  473.28 &  359.91 & 26571.42 & 10.0 : 3.7 : 2.8 \\
G3 & -0.72 & 26.56 & 4787.17 & 1680.64 & 1279.94 & 17397.76 & 10.0 : 3.5 : 2.7 \\
E1 &  1.95 & 23.82 & 1901.43 &  626.89 &  324.93 & 19170.60 & 10.0 : 3.3 : 1.7 \\
E2 &  0.97 & 26.43 &  986.60 &  356.65 &  260.88 & 15674.39 & 10.0 : 3.6 : 2.6 \\
E3 &  2.50 & 25.54 & 1045.67 &  378.20 &  294.89 & 15110.75 & 10.0 : 3.6 : 2.8 \\
P1 &  2.19 & 25.30 & 1810.98 &  609.40 &  478.93 & 16720.85 & 10.0 : 3.4 : 2.6 \\
P2 &  3.67 & 25.16 & 3513.34 & 1160.72 &  927.54 & 18069.53 & 10.0 : 3.3 : 2.6 \\
P3 & -0.84 & 26.06 & 1876.75 &  658.13 &  487.76 & 15228.11 & 10.0 : 3.5 : 2.6 \\

%% file: paramTexTableChangeR0.txt
   & 8.0 & 26.74 & 1525.19 &  569.35 &  382.73 & 14622.23  & 10.0 : 3.7 : 2.5 \\
G1 & 7.6 & 27.67 & 1430.76 &  528.21 &  363.45 & 14384.62  & 10.0 : 3.7 : 2.5 \\
   & 7.3 & 28.63 & 1357.09 &  494.16 &  349.04 & 14165.15  & 10.0 : 3.6 : 2.6 \\[10pt]
   & 8.0 & 25.91 & 1313.82 &  467.27 &  337.77 & 15676.58  & 10.0 : 3.6 : 2.6 \\
G2 & 7.6 & 26.65 & 1235.00 &  435.18 &  320.48 & 15303.12  & 10.0 : 3.5 : 2.6 \\
   & 7.3 & 27.42 & 1173.14 &  408.84 &  307.58 & 14993.34  & 10.0 : 3.5 : 2.6 \\[10pt]
   & 8.0 & 24.16 & 4587.68 & 1658.73 & 1330.48 &  9712.92  & 10.0 : 3.6 : 2.9 \\
G3 & 7.6 & 25.32 & 4266.68 & 1528.53 & 1272.76 &  9502.48  & 10.0 : 3.6 : 3.0 \\
   & 7.3 & 26.52 & 4011.04 & 1418.58 & 1226.61 &  9353.45  & 10.0 : 3.5 : 3.1 \\[10pt]
   & 8.0 & 21.41 & 1710.52 &  626.99 &  343.73 & 11530.12  & 10.0 : 3.7 : 2.0 \\
E1 & 7.6 & 22.48 & 1595.52 &  575.62 &  327.43 & 11028.20  & 10.0 : 3.6 : 2.1 \\
   & 7.3 & 23.70 & 1509.95 &  532.02 &  316.25 & 10707.93  & 10.0 : 3.5 : 2.1 \\[10pt]
   & 8.0 & 24.56 &  974.73 &  351.07 &  264.40 &  9135.87  & 10.0 : 3.6 : 2.7 \\
E2 & 7.6 & 25.63 &  907.63 &  323.75 &  250.86 &  9085.08  & 10.0 : 3.6 : 2.8 \\
   & 7.3 & 26.75 &  855.05 &  301.02 &  240.66 &  9043.39  & 10.0 : 3.5 : 2.8 \\[10pt]
   & 8.0 & 23.87 & 1023.09 &  365.74 &  297.10 &  9341.10  & 10.0 : 3.6 : 2.9 \\
E3 & 7.6 & 24.75 &  954.56 &  338.68 &  281.02 &  9254.02  & 10.0 : 3.5 : 2.9 \\
   & 7.3 & 25.69 &  900.61 &  316.34 &  268.93 &  9180.48  & 10.0 : 3.5 : 3.0 \\[10pt]
   & 8.0 & 22.05 & 1698.22 &  609.30 &  485.75 &  9418.38  & 10.0 : 3.6 : 2.9 \\
P1 & 7.6 & 23.34 & 1550.61 &  554.41 &  456.64 &  9288.41  & 10.0 : 3.6 : 2.9 \\
   & 7.3 & 24.68 & 1435.21 &  509.34 &  434.79 &  9191.40  & 10.0 : 3.5 : 3.0 \\[10pt]
   & 8.0 & 21.69 & 3192.12 & 1144.07 &  919.70 & 10148.09  & 10.0 : 3.6 : 2.9 \\
P2 & 7.6 & 23.03 & 2894.57 & 1035.28 &  861.54 &  9957.52  & 10.0 : 3.6 : 3.0 \\
   & 7.3 & 24.43 & 2663.48 &  946.72 &  817.88 &  9813.78  & 10.0 : 3.6 : 3.1 \\[10pt]
   & 8.0 & 23.61 & 1827.34 &  651.79 &  487.70 &  8606.41  & 10.0 : 3.6 : 2.7 \\
P3 & 7.6 & 24.77 & 1686.69 &  599.37 &  460.45 &  8566.67  & 10.0 : 3.6 : 2.7 \\
   & 7.3 & 25.88 & 1584.40 &  555.15 &  444.66 &  8479.57  & 10.0 : 3.5 : 2.8 \\[10pt]

%% file: bulge.bbl
\begin{thebibliography}{49}
\providecommand{\natexlab}[1]{#1}

\bibitem[{{Alves} et~al.(2002){Alves}, {Rejkuba}, {Minniti}
  et~al.}]{2002ApJ...573L..51A}
{Alves}, D.R., {Rejkuba}, M., {Minniti}, D., et~al., 2002, \apjl, 573, L51

\bibitem[{{Alves} \& {Sarajedini}(1999)}]{1999ApJ...511..225A}
{Alves}, D.R., {Sarajedini}, A., 1999, \apj, 511, 225

\bibitem[{{Babusiaux} \& {Gilmore}(2005)}]{2005MNRAS.358.1309B}
{Babusiaux}, C., {Gilmore}, G., 2005, \mnras, 358, 1309

\bibitem[{{Benjamin} et~al.(2005){Benjamin}, {Churchwell}, {Babler}
  et~al.}]{2005ApJ...630L.149B}
{Benjamin}, R.A., {Churchwell}, E., {Babler}, B.L., et~al., 2005, \apjl, 630,
  L149

\bibitem[{{Binney} et~al.(1991){Binney}, {Gerhard}, {Stark}
  et~al.}]{1991MNRAS.252..210B}
{Binney}, J., {Gerhard}, O.E., {Stark}, A.A., et~al., 1991, \mnras, 252, 210

\bibitem[{{Binney} \& {Tremaine}(1987)}]{1987gady.book.....B}
{Binney}, J., {Tremaine}, S., 1987, {Galactic dynamics}, Princeton, NJ,
  Princeton University Press, 1987, 747 p.

\bibitem[{{Bissantz} \& {Gerhard}(2002)}]{2002MNRAS.330..591B}
{Bissantz}, N., {Gerhard}, O., 2002, \mnras, 330, 591

\bibitem[{{Blitz} \& {Spergel}(1991)}]{1991ApJ...379..631B}
{Blitz}, L., {Spergel}, D.N., 1991, \apj, 379, 631

\bibitem[{{Bond} et~al.(2001){Bond}, {Abe}, {Dodd}
  et~al.}]{2001MNRAS.327..868B}
{Bond}, I.A., {Abe}, F., {Dodd}, R.J., et~al., 2001, \mnras, 327, 868

\bibitem[{{Bureau} et~al.(2006){Bureau}, {Aronica}, {Athanassoula}
  et~al.}]{2006MNRAS.370..753B}
{Bureau}, M., {Aronica}, G., {Athanassoula}, E., et~al., 2006, \mnras, 370, 753

\bibitem[{{Cabrera-Lavers} et~al.(2007){Cabrera-Lavers}, {Hammersley},
  {Gonzalez-Fernandez} et~al.}]{2007astro.ph..2109C}
{Cabrera-Lavers}, A., {Hammersley}, P.L., {Gonzalez-Fernandez}, C., et~al.,
  2007, ArXiv Astrophysics e-prints

\bibitem[{{Dwek} et~al.(1995){Dwek}, {Arendt}, {Hauser}
  et~al.}]{1995ApJ...445..716D}
{Dwek}, E., {Arendt}, R.G., {Hauser}, M.G., et~al., 1995, \apj, 445, 716

\bibitem[{{Dye} et~al.(2006){Dye}, {Warren}, {Hambly}
  et~al.}]{2006MNRAS.372.1227D}
{Dye}, S., {Warren}, S.J., {Hambly}, N.C., et~al., 2006, \mnras, 372, 1227

\bibitem[{{Efstathiou} et~al.(1988){Efstathiou}, {Ellis} \&
  {Peterson}}]{1988MNRAS.232..431E}
{Efstathiou}, G., {Ellis}, R.S., {Peterson}, B.A., 1988, \mnras, 232, 431

\bibitem[{{Eisenhauer} et~al.(2005){Eisenhauer}, {Genzel}, {Alexander}
  et~al.}]{2005ApJ...628..246E}
{Eisenhauer}, F., {Genzel}, R., {Alexander}, T., et~al., 2005, \apj, 628, 246

\bibitem[{{Evans} \& {Belokurov}(2002)}]{2002ApJ...567L.119E}
{Evans}, N.W., {Belokurov}, V., 2002, \apjl, 567, L119

\bibitem[{{Faria} et~al.(2007){Faria}, {Johnson}, {Ferguson}
  et~al.}]{2007AJ....133.1275F}
{Faria}, D., {Johnson}, R.A., {Ferguson}, A.M.N., et~al., 2007, \aj, 133, 1275

\bibitem[{{Gerhard}(2002)}]{2002ASPC..273...73G}
{Gerhard}, O., 2002, in G.S. {Da Costa}, H.~{Jerjen}, eds., ASP Conf. Ser. 273:
  The Dynamics, Structure \& History of Galaxies: A Workshop in Honour of
  Professor Ken Freeman, 73--+

\bibitem[{{Girardi} \& {Salaris}(2001)}]{2001MNRAS.323..109G}
{Girardi}, L., {Salaris}, M., 2001, \mnras, 323, 109

\bibitem[{{Ibata} \& {Gilmore}(1995)}]{1995MNRAS.275..605I}
{Ibata}, R.A., {Gilmore}, G.F., 1995, \mnras, 275, 605

\bibitem[{{Ishihara} \& {Onaka}(2006)}]{2006IAUJD..13E..11I}
{Ishihara}, D., {Onaka}, T., 2006, Exploiting Large Surveys for Galactic
  Astronomy, 26th meeting of the IAU, Joint Discussion 13, 22-23 August 2006,
  Prague, Czech Republic, JD13, \#11, 13

\bibitem[{{Lawrence} et~al.(2006){Lawrence}, {Warren}, {Almaini}
  et~al.}]{Law06}
{Lawrence}, A., {Warren}, S.J., {Almaini}, O., et~al., 2006, astro-ph/0604426

\bibitem[{{L{\'o}pez-Corredoira} et~al.(2000){L{\'o}pez-Corredoira},
  {Hammersley}, {Garz{\'o}n} et~al.}]{2000MNRAS.313..392L}
{L{\'o}pez-Corredoira}, M., {Hammersley}, P.L., {Garz{\'o}n}, F., et~al., 2000,
  \mnras, 313, 392

\bibitem[{{Lupton} et~al.(1987){Lupton}, {Gunn} \&
  {Griffin}}]{1987AJ.....93.1114L}
{Lupton}, R.H., {Gunn}, J.E., {Griffin}, R.F., 1987, \aj, 93, 1114

\bibitem[{{Minniti} et~al.(1995){Minniti}, {Olszewski}, {Liebert}
  et~al.}]{1995MNRAS.277.1293M}
{Minniti}, D., {Olszewski}, E.W., {Liebert}, J., et~al., 1995, \mnras, 277,
  1293

\bibitem[{{Nakada} et~al.(1991){Nakada}, {Onaka}, {Yamamura}
  et~al.}]{1991Natur.353..140N}
{Nakada}, Y., {Onaka}, T., {Yamamura}, I., et~al., 1991, \nat, 353, 140

\bibitem[{{Nikolaev} \& {Weinberg}(1997)}]{1997ApJ...487..885N}
{Nikolaev}, S., {Weinberg}, M.D., 1997, \apj, 487, 885

\bibitem[{{Nishiyama} et~al.(2005){Nishiyama}, {Nagata}, {Baba}
  et~al.}]{2005ApJ...621L.105N}
{Nishiyama}, S., {Nagata}, T., {Baba}, D., et~al., 2005, \apjl, 621, L105

\bibitem[{{Nishiyama} et~al.(2006){Nishiyama}, {Nagata}, {Sato}
  et~al.}]{2006ApJ...647.1093N}
{Nishiyama}, S., {Nagata}, T., {Sato}, S., et~al., 2006, \apj, 647, 1093

\bibitem[{{Omont} et~al.(2003){Omont}, {Gilmore}, {Alard}
  et~al.}]{2003A&A...403..975O}
{Omont}, A., {Gilmore}, G.F., {Alard}, C., et~al., 2003, \aap, 403, 975

\bibitem[{{Paczynski} \& {Stanek}(1998)}]{1998ApJ...494L.219P}
{Paczynski}, B., {Stanek}, K.Z., 1998, \apjl, 494, L219

\bibitem[{{Patsis} et~al.(2002){Patsis}, {Skokos} \&
  {Athanassoula}}]{2002MNRAS.337..578P}
{Patsis}, P.A., {Skokos}, C., {Athanassoula}, E., 2002, \mnras, 337, 578

\bibitem[{{Percival} \& {Salaris}(2003)}]{2003MNRAS.343..539P}
{Percival}, S.M., {Salaris}, M., 2003, \mnras, 343, 539

\bibitem[{{P\'{e}rez} et~al.(2006){P\'{e}rez}, {S\'{a}nchez-Bl\'{a}zquez} \&
  {Zurita}}]{Per06}
{P\'{e}rez}, I., {S\'{a}nchez-Bl\'{a}zquez}, P., {Zurita}, A., 2006,
  astro-ph/0612159

\bibitem[{{Rattenbury} et~al.(2006){Rattenbury}, {Mao}, {Debattista}
  et~al.}]{Rat06}
{Rattenbury}, N., {Mao}, S., {Debattista}, V., et~al., 2006, submitted to MNRAS

\bibitem[{{Salaris} et~al.(2003){Salaris}, {Percival}, {Brocato}
  et~al.}]{2003ApJ...588..801S}
{Salaris}, M., {Percival}, S., {Brocato}, E., et~al., 2003, \apj, 588, 801

\bibitem[{{Santiago} et~al.(2006){Santiago}, {Javiel} \& {Porto de
  Mello}}]{2006A&A...458..113S}
{Santiago}, B.X., {Javiel}, S.C., {Porto de Mello}, G.F., 2006, \aap, 458, 113

\bibitem[{{Skrutskie} et~al.(2006){Skrutskie}, {Cutri}, {Stiening}
  et~al.}]{2006AJ....131.1163S}
{Skrutskie}, M.F., {Cutri}, R.M., {Stiening}, R., et~al., 2006, \aj, 131, 1163

\bibitem[{{Stanek} et~al.(2000){Stanek}, {Kaluzny}, {Wysocka}
  et~al.}]{2000AcA....50..191S}
{Stanek}, K.Z., {Kaluzny}, J., {Wysocka}, A., et~al., 2000, Acta Astronomica,
  50, 191

\bibitem[{{Stanek} et~al.(1994){Stanek}, {Mateo}, {Udalski}
  et~al.}]{1994ApJ...429L..73S}
{Stanek}, K.Z., {Mateo}, M., {Udalski}, A., et~al., 1994, \apjl, 429, L73

\bibitem[{{Stanek} et~al.(1997){Stanek}, {Udalski}, {Szymanski}
  et~al.}]{1997ApJ...477..163S}
{Stanek}, K.Z., {Udalski}, A., {Szymanski}, M., et~al., 1997, \apj, 477, 163

\bibitem[{{Sumi}(2004)}]{2004MNRAS.349..193S}
{Sumi}, T., 2004, \mnras, 349, 193

\bibitem[{{Sumi} et~al.(2003){Sumi}, {Abe}, {Bond}
  et~al.}]{2003ApJ...591..204S}
{Sumi}, T., {Abe}, F., {Bond}, I.A., et~al., 2003, \apj, 591, 204

\bibitem[{{Sumi} et~al.(2004){Sumi}, {Wu}, {Udalski}
  et~al.}]{2004MNRAS.348.1439S}
{Sumi}, T., {Wu}, X., {Udalski}, A., et~al., 2004, \mnras, 348, 1439

\bibitem[{{Udalski}(2000)}]{2000ApJ...531L..25U}
{Udalski}, A., 2000, \apjl, 531, L25

\bibitem[{{Udalski} et~al.(1994){Udalski}, {Szymanski}, {Stanek}
  et~al.}]{1994AcA....44..165U}
{Udalski}, A., {Szymanski}, M., {Stanek}, K.Z., et~al., 1994, Acta Astronomica,
  44, 165

\bibitem[{{Udalski} et~al.(2000){Udalski}, {Zebrun}, {Szymanski}
  et~al.}]{2000AcA....50....1U}
{Udalski}, A., {Zebrun}, K., {Szymanski}, M., et~al., 2000, Acta Astronomica,
  50, 1

\bibitem[{{Weiland} et~al.(1994){Weiland}, {Arendt}, {Berriman}
  et~al.}]{1994ApJ...425L..81W}
{Weiland}, J.L., {Arendt}, R.G., {Berriman}, G.B., et~al., 1994, \apjl, 425,
  L81

\bibitem[{{Zhao} et~al.(2001){Zhao}, {Qiu} \& {Mao}}]{2001ApJ...551L..85Z}
{Zhao}, G., {Qiu}, H.M., {Mao}, S., 2001, \apjl, 551, L85

\end{thebibliography}
